%% file: 0_main.tex
\documentclass[sigconf]{acmart}

\AtBeginDocument{%
  \providecommand\BibTeX{{%
    \normalfont B\kern-0.5em{\scshape i\kern-0.25em b}\kern-0.8em\TeX}}}

\usepackage{acmart-taps}

\copyrightyear{2025}
\acmYear{2025}
\setcopyright{cc}
\setcctype{by}
\acmConference[CHI '25]{CHI Conference on Human Factors in Computing Systems}{April 26-May 1, 2025}{Yokohama, Japan}
\acmBooktitle{CHI Conference on Human Factors in Computing Systems (CHI '25), April 26-May 1, 2025, Yokohama, Japan}\acmDOI{10.1145/3706598.3714283}
\acmISBN{979-8-4007-1394-1/25/04}

\usepackage{subfig}
\usepackage{cleveref}
\usepackage{array}

\newif\ifcomment
\commentfalse

\newif\ifrevision
\revisionfalse 

\ifcomment

\ifrevision

\newcommand{\hide}[1]{}
\newcommand{\note}[1]{}
\newcommand{\added}[1]{\textcolor[rgb]{0, 0.094, 0.89}{#1}}
\newcommand{\modif}[1]{\textcolor[rgb]{0.5,0.5,0.9}{#1}}
\newcommand{\cut}[1]{}
\newcommand{\todo}[1]{}

\newcommand{\deleted}[1]{}

\newcommand{\temporary}[1]{}

\newcommand{\peisen}[1]{}
\newcommand{\christophe}[1]{}
\newcommand{\damien}[1]{}
\newcommand{\weitsang}[1]{}
\newcommand{\jeremie}[1]{}
\newcommand{\others}[1]{}

\else

\newcommand{\hide}[1]{}
\newcommand{\note}[1]{\textcolor{blue}{<< #1 >>}}
\newcommand{\added}[1]{\textcolor[rgb]{0.1, 0.56, 1}{#1}}
\newcommand{\modif}[1]{\textcolor[rgb]{0.5,0.5,0.9}{#1}}
\newcommand{\cut}[1]{\textcolor[rgb]{0.5,0.5,0.5}{CUT: #1}}

\newcommand{\todo}[1]{\textcolor{red}{TODO: #1}}

\newcommand{\deleted}[1]{\textcolor[rgb]{0.7,0.,0.}{{#1}}}

\newcommand{\temporary}[1]{\textcolor[rgb]{0.9,0.9,0.9}{{#1}}}

\newcommand{\peisen}[1]{\textcolor{blue}{Peisen: #1}}
\newcommand{\damien}[1]{\textcolor[rgb]{0.8,0.2,0.2}{Damien: #1}}
\newcommand{\jeremie}[1]{\textcolor[rgb]{0,0.7,0.8}{Jeremie: #1}}
\newcommand{\christophe}[1]{\textcolor[rgb]{0,0.8,0.4}{Christophe: #1}}
\newcommand{\weitsang}[1]{\textcolor[rgb]{0.8,0.8,0.0}{WeiTsang: #1}}
\newcommand{\others}[1]{\textcolor[rgb]{0.5,0.0,0.5}{#1}}

\fi

\else
\newcommand{\hide}[1]{}
\newcommand{\note}[1]{}
\newcommand{\added}[1]{#1}
\newcommand{\modif}[1]{#1}
\newcommand{\cut}[1]{}

\newcommand{\todo}[1]{}

\newcommand{\deleted}[1]{}

\newcommand{\temporary}[1]{}

\newcommand{\peisen}[1]{}
\newcommand{\christophe}[1]{}
\newcommand{\damien}[1]{}
\newcommand{\weitsang}[1]{}
\newcommand{\zsd}[1]{}
\newcommand{\jeremie}[1]{}
\newcommand{\others}[1]{}

\fi

\begin{document}

\title[SafeSpect]{SafeSpect: Safety-First Augmented Reality Heads-up Display for Drone Inspections}

\author{Peisen Xu}
\email{ps.xu@nus.edu.sg}
\orcid{0000-0003-1312-3061}
\affiliation{%
\institution{IPAL, National University of Singapore}
  \country{Singapore}
}

\author{Jérémie Garcia}
\email{jeremie.garcia@enac.fr}
\affiliation{%
  \institution{ENAC, Université de Toulouse}
  \city{Toulouse}
  \country{France}
}

\author{Wei Tsang Ooi}
\email{ooiwt@comp.nus.edu.sg}
\affiliation{%
  \institution{Computer Science}
  \institution{National University of Singapore}
  \country{Singapore}
}

\author{Christophe Jouffrais}
\email{christophe.jouffrais@cnrs.fr}
\affiliation{%
  \institution{CNRS, IPAL}
  \country{Singapore}
}
\affiliation{
  \institution{IRIT, Univ of Toulouse 3}
  \city{Toulouse}
  \country{France}
}

\renewcommand{\shortauthors}{Xu, et al.}



\input{Chapters/0_Abstract}




\begin{teaserfigure}
  \includegraphics[width=\textwidth]{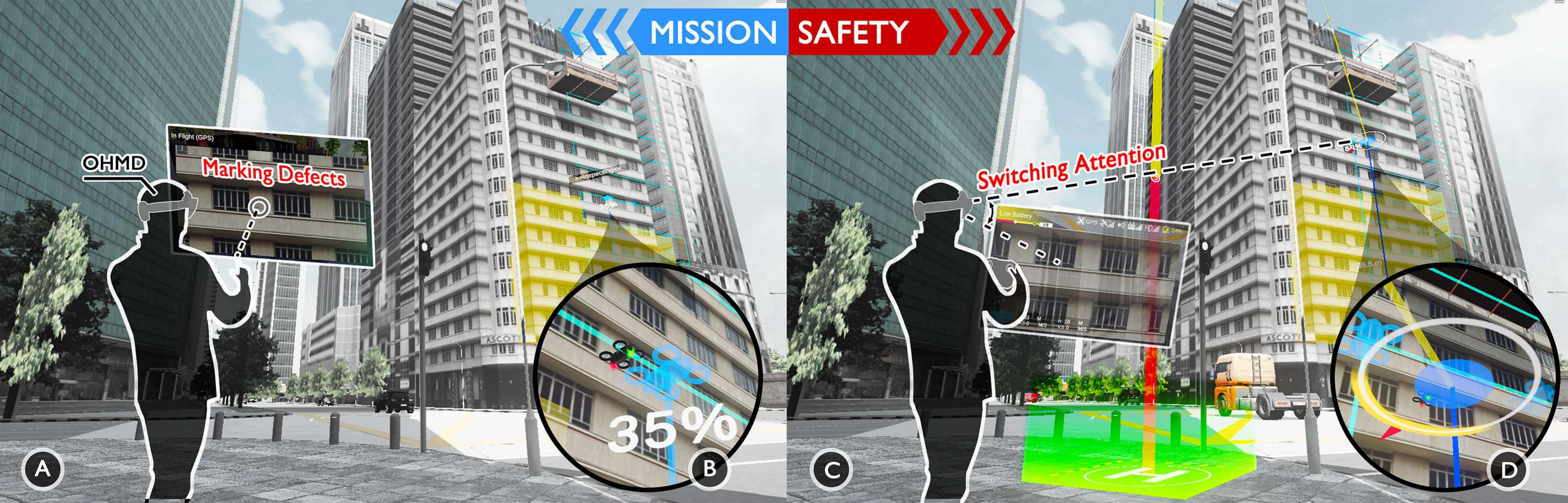}
  \caption{An adaptive AR drone control interface for building inspection task balances mission and safety information to reduce cognitive load. (A) shows a simplified interface for data collection. (B) helps track the drone's position and mission progress. (C) illustrates a safety view for attention switching between camera view and the line-of-sight. (D) displays the critical data such as heading, collision status and positional uncertainty near the drone for enhanced situational awareness.}
  \Description{This is the teaser figure that illustrates an adaptive AR drone control interface for building inspection task. The pilot is wearing an Optical Head-Mounted Display (OHMDs) for drone control and data collection. On the left, a mission view is illustrated which simplifies the interface to help the pilot focus on collecting data and monitoring the progress. Details of the AR waypoints and path visualization are shown. On the right, a safety view is displayed with flight data visualized in 2D and AR to make the pilot aware of the safety issues. This view helps the pilot switch attention between a 2D camera view and the direct line-of-sight for controlling the drone. The situational-awareness-enhancing AR visualizations, such as the drone heading, collision status, and positional uncertainty are shown.}
  \label{fig:teaser}
\end{teaserfigure}

\maketitle

\input{Chapters/1_Introduction}

\input{Chapters/2_Related_Work}

\input{Chapters/3_Design_Workshop}

\input{Chapters/4_System_Design}

\input{Chapters/5_Comparative_Study}
\input{Chapters/6_Discussion_Conclusion_Future_Work}

\begin{acks}

%

We are thankful for the support that this project has received from the National Research Foundation, Prime Minister’s Office, Singapore under its Campus for Research Excellence and Technological Enterprise (CREATE) programme. We would also like to thank all the pilots that accepted to share their time and ideas about their work, in particular, Gautier, Murat, Fabien, Xavier, Florian, Ashley, Wan Yi and Zheng Hao. We extend our gratitude to the members of Synteraction Lab for their
valuable feedback in different stages of our research.

\end{acks}

\bibliographystyle{ACM-Reference-Format}
\bibliography{1_reference}


\appendix
\input{Chapters/7_Appendix}

\end{document}

%% file: Chapters/0_Abstract.tex
\begin{abstract}

Current tablet-based interfaces for drone operations often impose a heavy cognitive load on pilots and reduce situational awareness by dividing attention between the video feed and the real world. To address these challenges, we designed a heads-up augmented reality (AR) interface that overlays in-situ information to support drone pilots in safety-critical tasks. Through participatory design workshops with professional pilots, we identified key features and developed an adaptive AR interface that dynamically switches between task and safety views to prevent information overload. We evaluated our prototype by creating a realistic building inspection task and comparing three interfaces: a 2D tablet, a static AR, and our adaptive AR design. A user study with 15 participants showed that the AR interface improved access to safety information, while the adaptive AR interface reduced cognitive load and enhanced situational awareness without compromising task performance. We offer design insights for developing safety-first heads-up AR interfaces.

\end{abstract}

%% file: Chapters/1_Introduction.tex
\section{Introduction}

Drone inspections have become commonplace for maintaining hard-to-reach areas in urban environments. 
In addition to structure condition monitoring, drones are used for data collection for digital twins and tracking construction progress~\cite{anwar2018construction, halder2023robots}. 
However, the current reliance on 2D handheld devices for flight control and data monitoring presents significant challenges for pilots \cite{whatMatters2021}. Pilots often experience "tunnel vision", focusing on the screen rather than the drone's physical surroundings, which increases the risk of collisions, especially in safety-critical scenarios \cite{borowik_mutable_2022}.
Head-up Displays (HUDs) and augmented reality (AR) interfaces provide an opportunity to integrate drone data directly into the pilot’s visual line of sight (VLOS), potentially reducing cognitive load \cite{ram2021lsvp, buchner_impact_2022} and improving situational awareness \cite{kim2009simulated}.  

A key challenge lies in designing an AR interface that effectively displays multiple critical elements such as remaining battery, GPS status, speed or orientation without overwhelming the pilot who is already facing cognitive burden~\cite{whatMatters2021}.
Previous research has often focused on individual elements of the interface, such as camera feed~\cite{hedayati2018improving}, waypoints (for semi-autonomous flight)~\cite{walker_robot_2019}, altitude~\cite{zollmann_flyar_2014} or heading~\cite{szafir2015communicating}. 
Rebensky \textit{et al.}~\cite{rebensky_impact_2022} suggest that directly adapting conventional 2D interfaces to heads-up displays might overwhelm new pilots, recommending AR HUDs to prioritize information based on task urgency and context.

Another challenge, as with every aeronautical activity, is to ensure safety. According to Rahmani et al.~\cite{rahmani_working_2023}, in addition to GPS signal loss or battery failures, situational awareness, decision-based, and skill-based errors are the main issues leading to drone accidents. While previous research has improved the operational efficiency of drone flight with AR, such as camera alignment~\cite{chen2021pinpointfly, hedayati2018improving}, less attention has been given to critical safety scenarios, which are difficult to simulate in real-world experiments.

Our goal is to address these two gaps: (1) the need to manage information processing during an inspection task with AR interfaces, without overwhelming the pilot, and (2) the lack of research on AR interfaces in safety-critical situations. We investigate the following research questions:

\begin{itemize}
    \item \textbf{RQ1}. How can 3D heads-up AR interfaces reduce the cognitive load of drone pilots performing facade inspection?
    \item \textbf{RQ2}. What design factors on the heads-up AR interface can ensure safety and situational awareness during critical scenarios of drone flight?
\end{itemize}

To answer these questions, we adopted a research-through-design approach \cite{RTD2007}, aiming at transforming drone interfaces \textit{"from their current state to a preferred state"} that enhances both safety and efficiency during drone inspections. Through this process, we sought to create an AR prototype that could improve drone piloting under challenging conditions while minimizing cognitive overload.

In this paper, we first describe the participatory design workshops with five professional pilots. These workshops helped identify the potential benefits of using AR for monitoring mission progress and ensuring safety during facade inspection, as well as the need to balance the information presented to avoid overwhelming users. We then describe the iterative design process with three pilots of a VR-based drone inspection simulator. During this process, we  explored how dynamic adaptation between safety and mission in AR could reduce cognitive load and enhance situational awareness. The Unity simulator allowed us to simulate the demanding inspection task and challenging scenarios such as GPS signal loss and obstacles without the risks associated with real-world testing. We then describe a user study with 15 experienced drone users comparing the adaptive AR interface (\textsc{adapt-ar}) with the conventional 2D interface (\textsc{2d-only}) and a static, non-adaptive AR interface (\textsc{full-ar}) during a facade inspection task. The study was followed by semi-structured interviews.
We conclude with design insights for developing safety-first heads-up AR interfaces.

%% file: Chapters/2_Related_Work.tex
\section{Related Work}

\subsection{Risks and Challenges of Urban Drone Flight}

Commercial drone operations, especially those conducted in densely populated urban areas, require rigorous safety assessment procedures. Through interviews with professional pilots, Ljungblad \textit{et al.}~\cite{whatMatters2021} identified safety as the foremost concern of the pilots. Rakotonarivo \textit{et al.}~\cite{rakotonarivo2023cleared} discovered three key factors that impact flight safety: (1) environmental factors like weather condition, geographical elements, airspace characteristics; (2) drone-specific parameters like automation, range, speed, and (3) mission-specific parameters such as payload, trajectories, timing of flight, take-off and landing zone. An online survey~\cite{rahmani_working_2023} highlighted situational awareness, decision-based errors, and skill-based errors as the primary human-factor-related causes of drone accidents. Poor communication, information display, and ambiguity in control modes were also identified as significant obstacles to effective human-drone collaboration. Moreover, Borowik \textit{et al.}~\cite{borowik_mutable_2022} showed with eye trackers that both beginner and advanced pilots tend to focus heavily on the camera feed during flights, leading to tunnel vision and near-collisions. 

\subsection{Heads-up AR Drone Control and Situational Awareness Support}

Following the distinctions outlined by Meta of AR, MR, and VR~\cite{Meta_2024}, we focus on AR that overlays digital elements onto the physical world, enhancing the user's perception, rather than MR, which focuses on the interaction between real and virtual content. 

Situational awareness (SA) was introduced by Endsley \textit{et al.}~\cite{endsley2000situation} as the understanding of what is happening around us, the current situation and the one to come. AR has been shown to improve situational awareness by visualizing the 2D camera feed~\cite{hedayati2018improving} and the signal status~\cite{rebensky_impact_2022} within the visual line-of-sight (VLOS) compared to a heads-down interface for drone operations. Addin \textit{et al.}~\cite{addin2021design} compared a AR heads-up interface with a PC interface during critical drone flight operations. Their results suggest that augmented reality can provide significant situational awareness support for those involved in stressful situations.

In addition to showing flight parameters on a heads-up display, Garcia~\textit{et al.}~\cite{garcia:hal-02128390} helps the safety pilots locate the drone with a localization ring. Previous work also showed that incorporating 3D AR visual elements, such as in-situ depictions of the drone's next movements~\cite{walker2018communicating, walker_robot_2019} or visualizing flight-path~\cite{chen2021pinpointfly} enhances operational efficiency, accuracy and usability of the interface. Similarly, visualizing spatial relations between the drone and its surroundings through 3D projections and path connections aids depth perception~\cite{zollmann_flyar_2014}. Alternatively, Inonue \textit{et al.}~\cite{inoue2023birdviewar} visualize the height and the proximity of the drone via an augmented Third Person View (TPV) to enhance the perception of the drone's surroundings. Nevertheless, their setup requires a follower drone to observe and control the main drone. 

Most of these studies were conducted either in indoor environments or outdoors in open fields and did not consider safety factors like GPS loss or wind turbulence. Additionally, many prior studies focused either on a limited set of information, such as the camera feed~\cite{hedayati2018improving} and waypoints~\cite{chen2021pinpointfly, zollmann_flyar_2014} or on only one modality of information (i.e. 2D~\cite{rebensky_impact_2022} or 3D~\cite{zollmann_flyar_2014}).
Rebensky \textit{et al.}~\cite{rebensky_impact_2022} reported that presenting all flight data and the camera feed in 2D on a heads-up display can lead to visual clutter and increased cognitive load, which motivates further investigation into effective information presentation on AR interfaces.

\subsection{Cognitive Load Management in AR}

\modif{Ryo \textit{et al.}~\cite{10.1145/3491102.3517719} synthesized a taxonomy of AR and robotics, identifying real-time data visualization and explainable robotics via AR as key research avenues. Buchner \textit{et al.}~\cite{buchner_impact_2022} reviewed AR applications, finding that AR generally reduces cognitive load by addressing the split-attention effect~\cite{Ayres2012}, which arises when information is spread across multiple sources. 
This aligns with the Cognitive Theory of Multimedia Learning~\cite{Mayer_2009}, where spatial and temporal contiguity—such as real-time feedback within the visual line of sight (VLOS)—helps minimize extraneous cognitive load. 
Further, Li \textit{et al.}~\cite{li2023comparative} showed that 3D AR interfaces improve eye fixation patterns and reduce perceived workload compared to 2D AR. 
Despite these advances, Buchner \textit{et al.}~\cite{buchner_impact_2022} highlighted a need for studies on value-added AR design principles. Our work addresses this gap by investigating how heads-up AR interfaces can reduce cognitive load in drone control tasks.}

%

\modif{In addition, adaptive AR interfaces have been explored to mitigate information overload. Lindlbauer \textit{et al.}~\cite{10.1145/3332165.3347945} emphasized the importance of adapting AR interfaces to user context, while Tatzgern \textit{et al.}~\cite{7504691} introduced a level-of-detail (LOD) technique using hierarchical clustering to organize AR elements. Lu \textit{et al.}~\cite{10.1145/3491102.3517723} demonstrated that semi-automated adaptive systems improve task performance. However, these efforts have not addressed safety-critical scenarios like drone control, where managing cognitive load and ensuring situational awareness are paramount. Our research extends this work by applying adaptive AR design principles to enhance drone control interfaces in such contexts.}

%% file: Chapters/3_Design_Workshop.tex
\section{Participatory Design Workshops}

We conducted a study with professional pilots to better understand the practical workflow and challenges of drone-based building inspections and identify the design requirements for a heads-up interface. We used participatory design techniques~\cite{schuler1993participatory} to generate and validate design concepts.
\subsection{Method}

\subsubsection{Participants}
We recruited five expert pilots from the French Civil Aviation School, all of whom hold a professional certificate for conducting drone flights and have more than 3 years of relevant experience in the industry. The participants are currently working in the research field of drone development. The details of the pilots are listed in Table~\ref{tab:design-workshop-participants} (D1-D5). 

We organized two separate sessions of the design workshop to accommodate all participants’ schedules. The first session involved three pilots (D1-D3) and three researchers from the group as facilitators and design assistants. The second session involved two pilots (D4, D5) and two researchers.

\begin{table*}[h]
  \caption{Professional Pilots Involved in the Participatory Design Workshop (D1-D5) and the Iterative Design (D5-D7)}
  \label{tab:design-workshop-participants}
  \begin{tabular}{ m{1em}  m{3em} m{2em}  m{34em}  m{8em} }
    \toprule
    ID&Gender&Age&Experiences&Session\\
    \midrule
    D1 & Male & 30 & 6 years of experience in drone development, mainly in embedded system design and operating ground control station (GCS) & Workshop 1 \\
    D2 & Male & 56 & 15 years of drone development with test flight for fixed wings and rotorcrafts. & Workshop 1 \\
    D3 & Male & 25 & 1 year of drone pilot, 3 years of drone research. Worked with home-made drones and designed drone control algorithms & Workshop 1 \\
    D4 & Male & 42 & 18 years of drone development and experiment with fixed wings and rotorcraft & Workshop 2 \\
    D5 & Male & 42 & 15 years of drone development and research. Conducted flight for scientific research with prototypes and some commercial drones. & Workshop 2 \& \textit{Iterative Design}\\
    \midrule
    D6 & Male & 28 & 3 years working experience in commercial aerial filming & \textit{Iterative Design} \\
    D7 & Male & 31 & 4 years of experience as drone instructor and 2 years experience in facade inspection & \textit{Iterative Design} \\
  \bottomrule
\end{tabular}
  \Description{Table detailing the ID, Gender, Age, Experience, and Session attended for our participants.}
\end{table*}

\subsubsection{Procedure and Apparatus}

\modif{After a brief introduction regarding the goals of the workshop, we asked the pilots to brainstorm on how they would conduct the building inspection mission and what were the challenges during the mission. 
We provided a printed photo of a high-rise building in urban area, taken from a pedestrian point of view. We also printed illustrations of essential data for drone control extracted from the DJI GO 4~\cite{DJI_GO_4} and DJI GS Pro~\cite{DJI_2018} interfaces that includes the typical flight data such as signal strength, flight mode, map actions, flight telemetry, way-points, obstacle detection, and avoidance.
Before moving to the design phase, participants tried a specific HoloLens AR application to familiarize themselves with the technology. They could use gestures to control a virtual drone's take-off, landing, and autopilot.}
%
\modif{Each pilot was paired with a researcher who would assist in generating design ideas.}
They were asked to imagine conducting the mission with a AR headset and illustrate the interface that would be helpful for the mission and addressing the challenges. 
We used transparent film papers to draw the design ideas simulating AR views.
After that, each pilot took turns sharing their ideas and drawings to trigger discussions.

\subsection{Results}

\modif{Participants actively engaged in the workshop, providing valuable insights into drone-based building inspections and contributing to the development of heads-up AR concepts for drone operations. Using low-fidelity prototypes, we explored AR solutions addressing three main areas: safety information, mission tracking, and adaptive visualizations to balance safety and mission-related information. Below, we detail our findings related to mission planning, operational challenges, and proposed AR design ideas.}

	
\subsubsection{Mission Planning}

\modif{Participants discussed inspecting a high-rise facade in a densely populated area, emphasizing safety, battery optimization, and photo coverage. D1 highlighted the need for safe landing zones and proposed starting inspections from the top to ensure safe battery management, though starting at the base could improve visibility for larger structures. D2 stressed the importance of photo overlap for photogrammetry, even at the cost of higher energy consumption. To address the building’s curved geometry, participants recommended consistent waypoint distances, with collision detection aiding safe piloting. D5 emphasized defining flight zones to prevent drone drift.}





\subsubsection{Challenges}
\paragraph{Split Attention} 

\modif{All participants explained that for legal reasons, pilots must maintain a visual line of sight (VLOS) with the drone, but they also had to rely on the camera view for navigation and identifying defects. This becomes challenging when the drone is distant, making it hard to track its location and orientation (D1, D2). D1 stressed the need to frequently look up at the drone because the camera view alone is insufficient for assessing safety. D2 noted that VLOS piloting improves situational awareness since \textit{“with the camera view, you cannot move left or right quickly; you have to turn first.”} However, managing piloting with safety signal monitoring and data acquisition simultaneously causes information overload, often requiring to operators (D3).}

D5 introduced the idea of ``autonomy level,'' suggesting that with an unsupervised flight mode --- where the drone follows a predefined path, maintains a set distance, and avoids obstacles --- a single pilot could manage both flight and data acquisition. He, however, doubted that he could program it with the current 2D tablet interface.

\paragraph{Battery Life}	

Pilots were concerned with battery life, often proposing solutions such as battery replacement stations on balconies or landing to swap batteries. Pilots rely on low-battery alarms to interrupt missions (D1, D3). However, resuming missions to the latest observed point is difficult as they usually lack clear visual cues. D1 suggested recording the GPS location of flight interruptions. D1 proposed showing an estimate of remaining flight time to reduce stress and optimize mission planning.

\paragraph{GPS loss and Drifting Issues} 

\modif{GPS loss and drifting are critical safety challenges that require immediate manual control (D1, D3). 
D1 emphasized that \textit{“This is the situation when you have to look at the drone. If you don’t know the heading, you don’t know what command you should use.”}.
Most commercial drones can handle moderate wind drifting without landing (D2). However, wind turbulence or GPS loss makes self-balancing unreliable, requiring manual altitude adjustment (D1, D2). }

\subsubsection{Heads-up Display Design Ideas}

Building on the discussed mission challenges, we used paper prototyping techniques to create and discuss AR visualizations related to the challenges identified in the scenario discussion phase. We detail our findings covering three main goals: enhancing safety, improving usability, and balancing information load.

\paragraph{Use AR to Enhance Safety}

\begin{figure}
    \centering
    \includegraphics[width=1\linewidth]{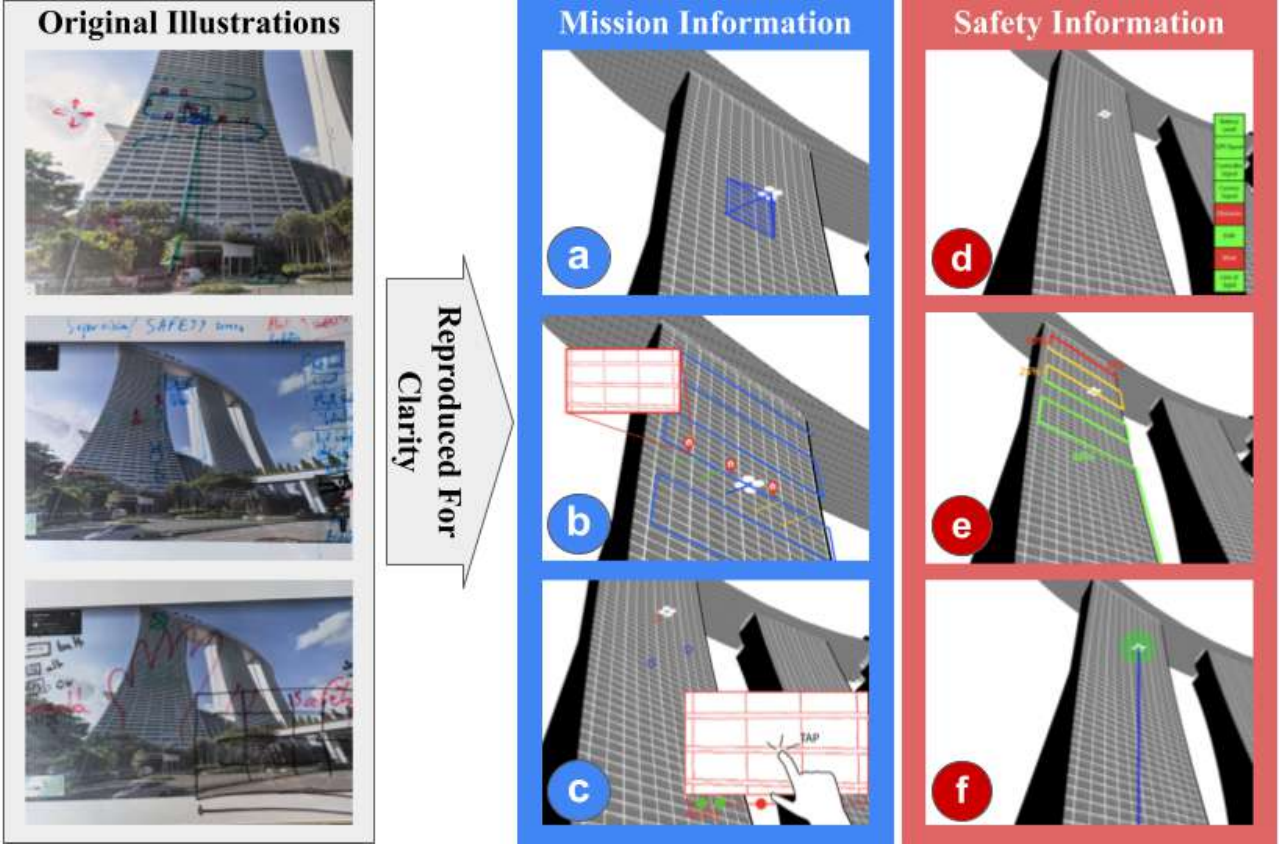}
    \caption{Examples of the safety and mission-related visualization illustrated by the pilots. (a) Display the heading of the drone, illustrated by D3. (b) Show the nearby waypoints and the photo-taking points over the trajectory, illustrated by D1. (c) Allow marking the point of interest from the camera view while monitoring, illustrated by D3. (d) Visualize safety signals at the peripheral vision, illustrated by D2. (e) Show the RTH path with battery remaining energy, illustrated by D1. (f) Display a boundary around the drone and the ground projection, illustrated by D4.}
    \label{fig:design-ideas}
    \Description{This figure shows examples of the safety and mission-related visualization illustrated by the pilots on the left and reproduced version on the right including (a): Display the heading of the drone, illustrated by D3 (b): Show the nearby waypoints and the photo-taking points along the trajectory, illustrated by D1 (c): Allow marking the point of interests from the camera view while monitoring, illustrated by D3. (d): Visualize safety signals at the peripheral vision, illustrated by D2. (e): Show the RTH path with battery remaining energy, illustrated by D1. (f): Display a boundary around the drone and the ground projection, illustrated by D4.}
\end{figure}

Most pilots mentioned the importance of showing the heading of the drone in the AR view, which helps the pilot know the orientation during manual piloting as presented in Figure~\ref{fig:design-ideas}.a.

Pilots also suggested various solutions for GPS and wind drifting issues. D1 proposed visualizing wind turbulence and GPS-denied areas as danger zones, but others found it counterproductive if it blocked the drone's view. D3 suggested displaying real-time wind direction and intensity above the drone, while D2 focused on using peripheral signals to indicate safety status as illustrated in Figure~\ref{fig:design-ideas}.d. 

D4 proposed to visualize a ground projection from the current position of the drone, indicating the horizontal positioning and the ground area influenced by the fall damage. It can help the pilot estimate the ground risk from the current midair position. To mitigate the danger of midair collision, a spherical boundary can be visualized around the drone for an intuitive perception of the safety distance when piloting with VLOS. The color of the boundary is dependent on the current nearest obstacle distance as detected by the drone (D4, Figure~\ref{fig:design-ideas}.f). Alternatively, D1 proposed visualizing a projection casting from the drone to the surface of the building to indicate the distance (Figure~\ref{fig:design-ideas}.b). 

D2 drew a path connecting the current drone location and the landing point. In addition, D1 and D5 brought up the idea of showing the needed battery remaining energy over this landing path, so that the pilot can glance at the visualization to know whether they have sufficient time to complete the mission (Figure~\ref{fig:design-ideas}.e). It helps the pilot project into a future state of the mission and prepares for landing, reducing the negative feeling of bewilderment when the battery alert occurs and improving situational awareness.

\paragraph{Use AR to Track Mission and Improve Usability}

All pilots agreed on the value of using AR to visualize waypoints and the flight trajectory over the building surface. D1 suggested displaying only nearby waypoints, while D3 recommended coloring completed segments of the trajectory differently. D1 and D5 added that marking the locations where photos were taken along the path would help track progress (Figure~\ref{fig:design-ideas}.b). D5 also highlighted that visualizing coverage across the building surface could assist in monitoring data quality and revisiting poorly captured points.

Several pilots stressed the benefit of integrating the camera feed within the VLOS view, allowing smoother attention shifts between tracking mission progress and focusing on the inspection task. D3 emphasized the need to mark points of interest (POIs) during the flight, such as areas with defects, low lighting, or locations where the drone may need to return home due to a low battery. He proposed using a simple tap gesture on the camera feed to mark defective regions for future reference (Figure~\ref{fig:design-ideas}.c).

\paragraph{Balancing Mission and Safety Information to Reduce Complexity}

Pilots expressed concerns about the complexity of the AR visualization that could occlude some elements of the real world or create information overload (D2, D3). To address this, they proposed separate views for mission and safety to streamline information. The mission-focused view would include the camera feed, sensor data (eg. infrared sensors), waypoints, camera frustum, and coverage with marked points of interest, along with a status message for warning notifications. The safety-focused view would prioritize safety data monitoring, including GPS satellite count, connectivity, battery levels, wind turbulence, and obstacle detection using high-contrast colors like red for urgent signals and green for normal ones as illustrated in Figure~\ref{fig:design-ideas}.d.

Additionally, pilots suggested adapting the camera feed's placement depending on the context (D1, D4). During autopilot, the camera feed should be centered to focus on data collection. When manual piloting is required due to safety issues, the camera feed should be positioned peripherally, allowing pilots to use AR overlays (collision boundaries, headings, RTH paths) for better situational awareness while managing both the feed and VLOS.

%% file: Chapters/4_System_Design.tex
\section{Iterative Design of an Adaptive AR Interface and a Realistic Inspection Task}\label{sec:system-design}

Using a technology probe approach~\cite{hutchinson2003technology} with three professional drone pilots (see table~\ref{tab:design-workshop-participants}, D5-D7), we iteratively developed an interactive prototype and a challenging inspection task.
We further explored how an AR interface could support pilots in safety-critical conditions and focus on the adaptation between safety and mission tasks. To address technical and safety challenges, we created a simulator in Unity \added{(Version 2022.3.12f1)} and tested it in a VR environment with a Meta Quest 3 headset. This setup allowed us to focus on AR design without the risks of real-world testing. 
Pilots provided insights on AR views, system usability, and task realism, as well as on how to improve the interface and adapt it to various operational scenarios.
We describe below the simulation environment, inspection task, and how we integrated safety challenges, followed by the interactions and visualizations that support pilots in managing their missions and responding to events.

\subsection{Inspection Task and Drone Simulation}\label{task-simulation}

\begin{figure}
    \centering
    \includegraphics[width=1\linewidth]{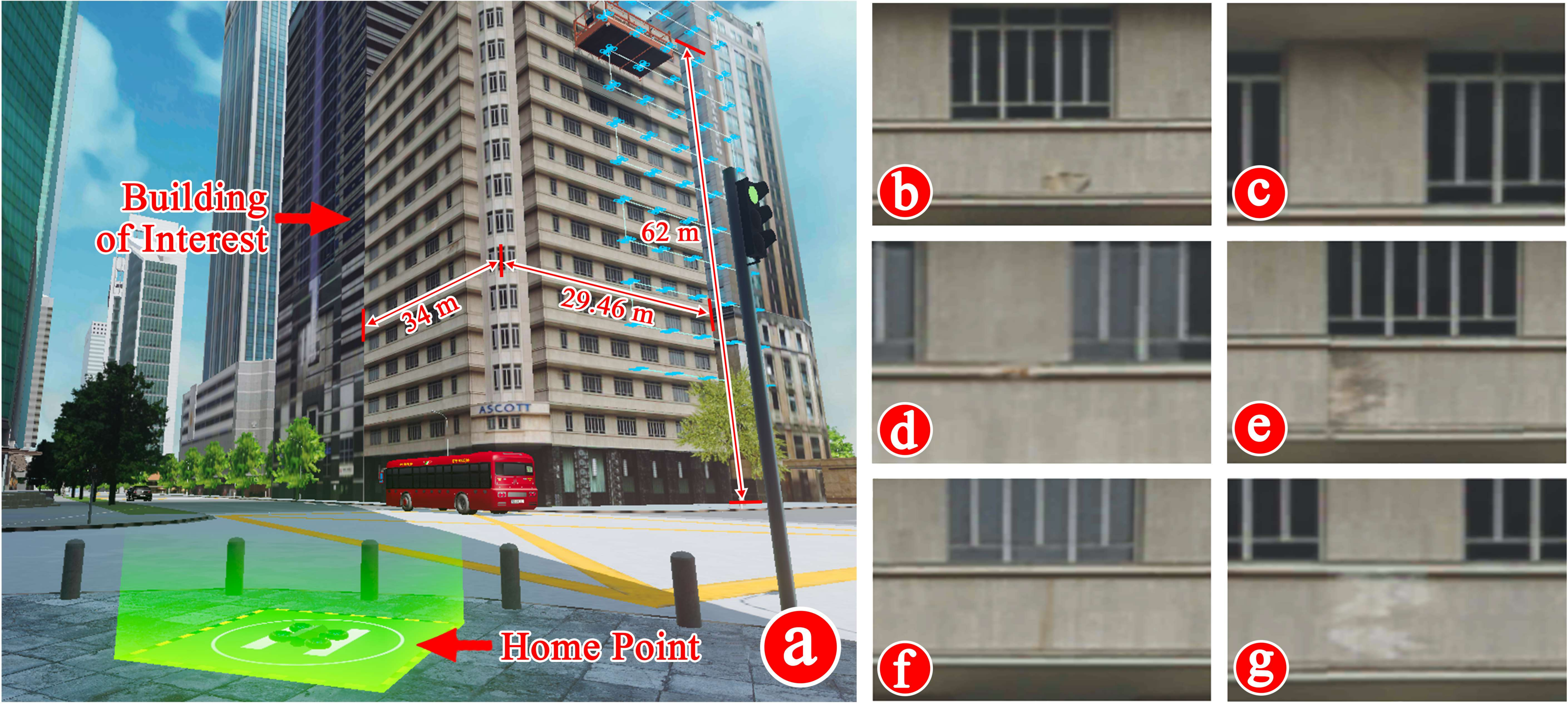}
    \caption{The simulated building inspection scenario and different types of defects used in our study. (a) Overall configuration of the scenario. (b) Paint peel. (c) Wall crack. (d) Wall dent. (e) Rotten surface. (f) Leakage. (g) Delamination. }
    \label{fig:scenario}
    \Description{This figure is a screenshot from our prototype application containing the location of the drone, a street, and a building to be inspected on the left. On the right, it shows examples of defects that are used in our study over the facade.}
\end{figure}

To simulate the building inspection task, we built a 3D environment based on the city center of Singapore as illustrated in Figure~\ref{fig:scenario}. The dimensions of the inspected building were 34m (width) x 29.46m (length) x 62m (height). It is located at the corner of the street and the standing location of the pilot is on the opposite corner. We simulated the busy traffic with ambient noises of the street to add realism. 

To create a realistic virtual drone, we adapted the velocity-based drone controller from Belkhale \textit{et al.}~\cite{UnityDroneSim}. Based on pilots' feedback, we finely tuned its behaviors to ensure its flight mechanics closely resemble those of a real one. We also created a drone simulation system that allows path planning, switching between manual and autopilot, battery, GPS, and wind simulation, and interactive visualization on 2D and AR interfaces. 

The mission objective is to correctly spot all defects on the building facade. We rendered six types of defects on the building facade, based on the common defects classifications shown in~\cite{lee_multidefectnet_2020, BCA_2022}. These include paint peel, wall crack, wall dent, rotten surface, leakage, and delamination (Figure~\ref{fig:scenario}.b-g). To avoid repetition, we created at least three variations for each type of defect. \modif{We intentionally made some defects hard to spot by placing them under the shadowed area on the building surface (Figure~\ref{fig:scenario}.c).} The defects are evenly laid out over the building facade, with 0-2 defects per layer. There are in total 11--12 defects on the entire facade of 12 layers. D7, who had experience with building inspection, found the defects realistic for the simulated task.


\subsection{Implementation of the Camera View and Interactions}

\begin{figure}
    \centering
    \includegraphics[width=1\linewidth]{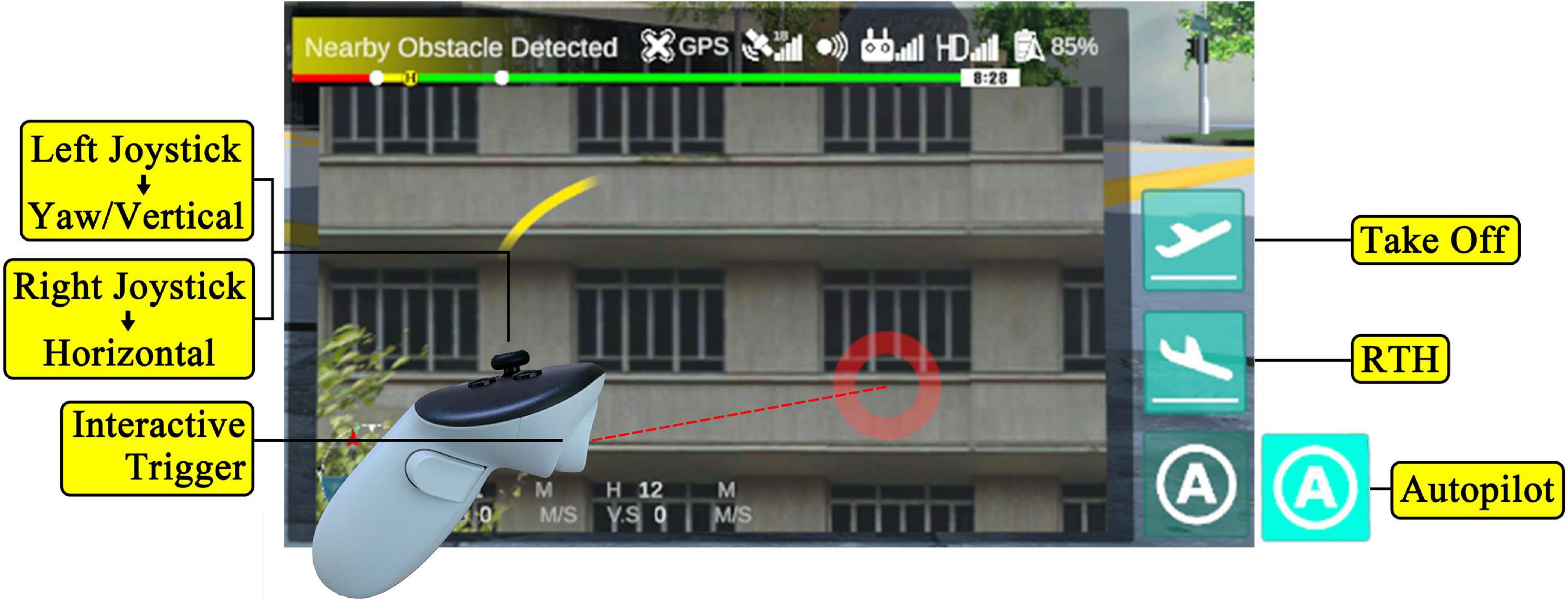}
    \caption{Interactive 2D Camera View. Left: Inputs from the Meta Quest Touch Pro VR Controllers (Only the Left Controller is shown in the picture). Right: Interactive buttons on the interface. The autopilot button (bottom right) will be highlighted if the autopilot is on. }
    \label{fig:interaction}
    \Description{This figure shows a screenshot of the video feed from our prototype, how the remote controller can be used to trigger interactive buttons or video capture.}
\end{figure}

Following suggestions from our participatory design workshops, we created a floating midair camera view in AR. In addition to the drone video feed, usual flight information is displayed on the camera view as illustrated in Figure~\ref{fig:interaction}. We adopted the user interface design of DJI GO 4~\cite{DJI_GO_4}, a popular flight control software developed by DJI, which features flight status messages, signal and system monitoring, a battery slider bar, telemetry data monitoring, and collision detection status. This implementation was based on suggestions by the pilots, saying that it would be a good option to rely on the familiar layout of the existing 2D interface, even in the new AR adaptive interface. 

The user interacts with the simulated drone via the 2D interface and the joysticks of the controllers. There are three buttons to manage the flow of the drone flight: a take-off button, a return-to-home (RTH) button, and an autopilot toggle button. The autopilot function can only be used when near the trajectory, and it can be interrupted by any manual input from the joysticks. The user can trigger the defect marking by directly "clicking" on the camera view, which creates a red circle on the exact location of the marked point (Figure~\ref{fig:interaction}). 

\subsection{Simulating Safety Challenges}

We simulated GPS loss, wind turbulence, and limited battery life as safety challenges that were identified during the workshops to replicate complex situations and design effective adaptation behaviors for such situations. Following the pilots' suggestions (D5-D7), we also added mid-air obstacles. Both GPS loss and obstacle detection cause the autopilot to stop, imposing pilots to manually operate the drone before enabling the autopilot again when safety issues are solved.

\subsubsection{GPS Loss and Drifting}

The GPS loss often occurs when the drone is close to the building. We manually defined several GPS-denied zones on the building facade. The GPS signal gradually decreases when entering the zone and gradually recovers when leaving the zone. We first implemented sudden changes but D5 commented that this did not feel realistic and D7 argued that \textit{"Normally the drone can still stabilise to a certain extent with visual sensors. Thus, the drift of global positioning should be a gradual process."}

When the GPS is lost, the system cannot confidently determine the accurate global position of the drone, resulting in inaccuracies of drone-centered visualizations. This prevents the drone from stabilizing using positional data, which makes it start drifting due to the wind. In our simulation, the wind is represented as a force vector that gradually adjusts its direction and intensity to match a target force vector, which is updated at random intervals. The behavior was adjusted from the previous iteration, using constant direction based on D7's feedback that \textit{"one difficulty about flying in real world is that the wind isn't constant. It can change direction too."}) The wind vector imposes force on the drone as long as it has taken off and affects the movement of the drone depending on the GPS status.


\begin{figure}[ht]
    \centering
    \includegraphics[width=0.9\linewidth]{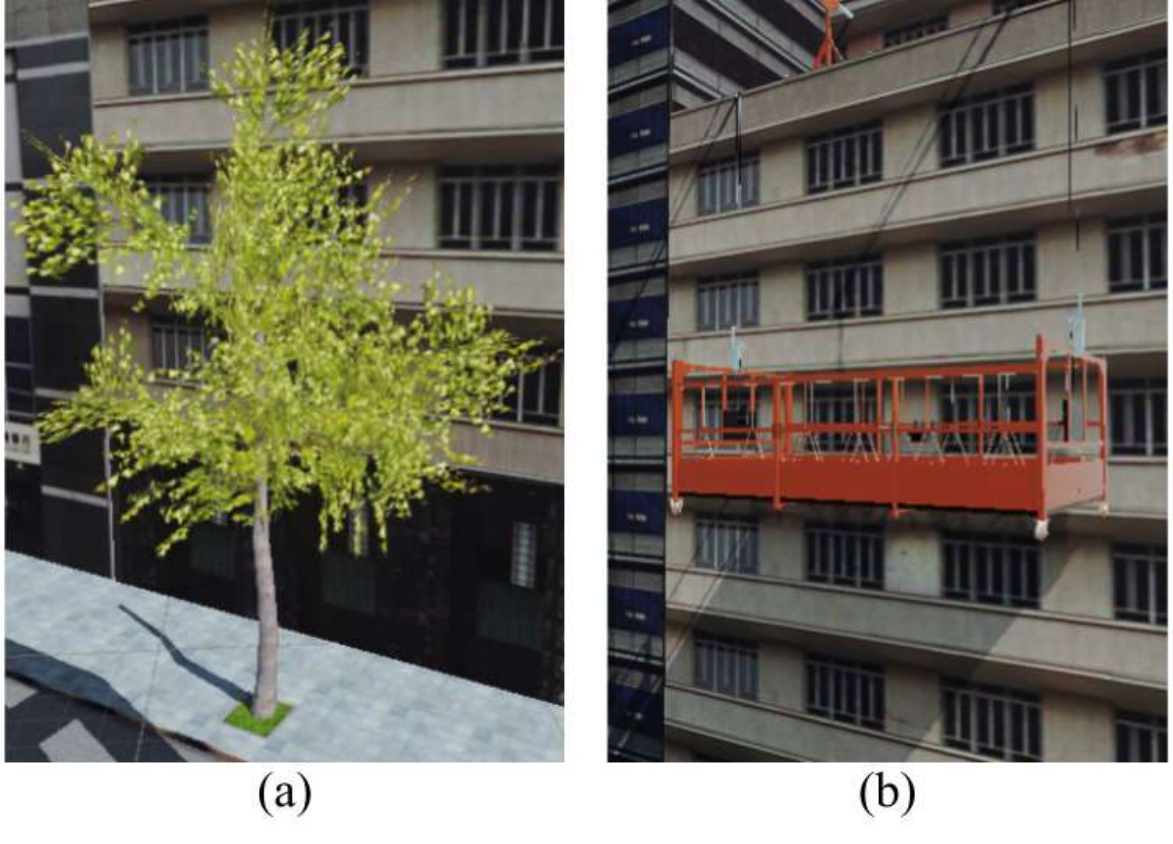}
    \caption{Mid-air obstacles. (a) A tall tree. (b) A gondola machine.}
    \label{fig:obstacles}
     \Description{This figure shows two screenshots of the simulated environment featuring (a) a tall tree in front of the facade and (b) a gondola suspended from the roof.}
\end{figure}

\subsubsection{Mid-air Obstacles}
Commenting on the difficulty of our task, D7 explained that \textit{"there are way more things to consider in real world flight"} such as poles extruding from the building, open windows, etc. The presence of unexpected mid-air obstacles, especially those blocking the flight path, imposes great challenges for drone navigation. We thus modelled two types of mid-air obstacles over the facade: a tall tree and a gondola machine in operation, as illustrated in Figure \ref{fig:obstacles}. Typically, the drone cannot reliably detect small branches on a tree or thin wires with its sensor, forcing pilots to rely on direct sight to avoid obstacles. 


\subsection{Mission and Safety AR Elements}

Mission and safety information are displayed as 3D in-situ visual cues, either centered on and moving with the drone or anchored in a fixed world position. 
The 2D panel presenting the video feed from the drone and safety information is anchored either to a head-fixed or a body-fixed location, depending on the mission context.
We describe our AR designs following the main flight phases including pre-inspection, inspection, critical events, and returning to home. 


\subsubsection{Pre-inspection and Inspection Phases}

\begin{figure}
    \centering
    \includegraphics[width=1\linewidth]{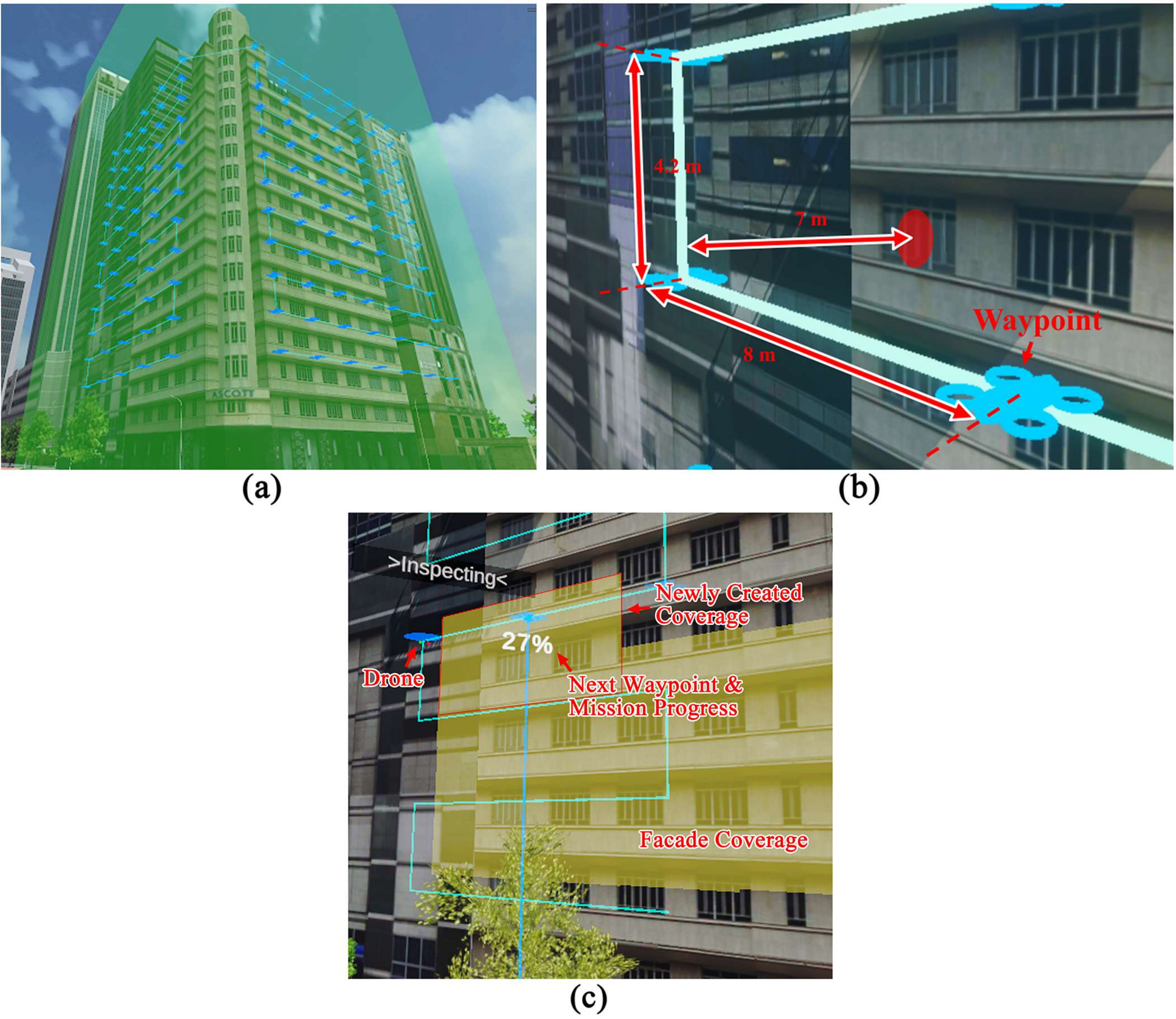}
    \caption{(a) Mission boundary (in green). (b) Generated path plan (in blue) based on building geometry and parameters. (c) Waypoints and coverage visualization during autpilot.}
    \label{fig:path-plan}
    \Description{This figure illustrates the AR holographic views of the drone at the planned waypoints and the lines between them on the left. Details about the generated flight plan including specific distances to the wall are given. }
\end{figure}

At the start of the mission, a safety buffer is visualized as a transparent green box surrounding the building, marking the safe area for operation (Figure~\ref{fig:path-plan}.a). A 3D path plan is projected onto the building facade, with each waypoint represented by a holographic drone model, as shown in Figure~\ref{fig:path-plan}.b.c.

Once the drone enters the mission boundary, the safety buffer visualization disappears, the autopilot can be activated and only the nearest waypoint to the drone's real-time position is shown, helping the pilot select the starting point for autopilot. 

During autopilot, only the next waypoint is displayed, following a suggestion from D1 during the workshop. Also, the drone travels at a constant speed between waypoints, pausing briefly upon reaching each one. The pauses allow the pilot to mark defects without the risk of a camera shake, improving the quality of data collection. Coverage visualizations appear on the facade at each waypoint, confirming that data has been successfully captured for that region (Figure~\ref{fig:path-plan}.c). The coverage visualization can also be displayed during the manual flight if the pilot briefly holds the drone's position at the waypoint. 



\subsubsection{Critical and Returning Phases}

\begin{figure}[!h]
    \centering
    \includegraphics[width=1\linewidth]{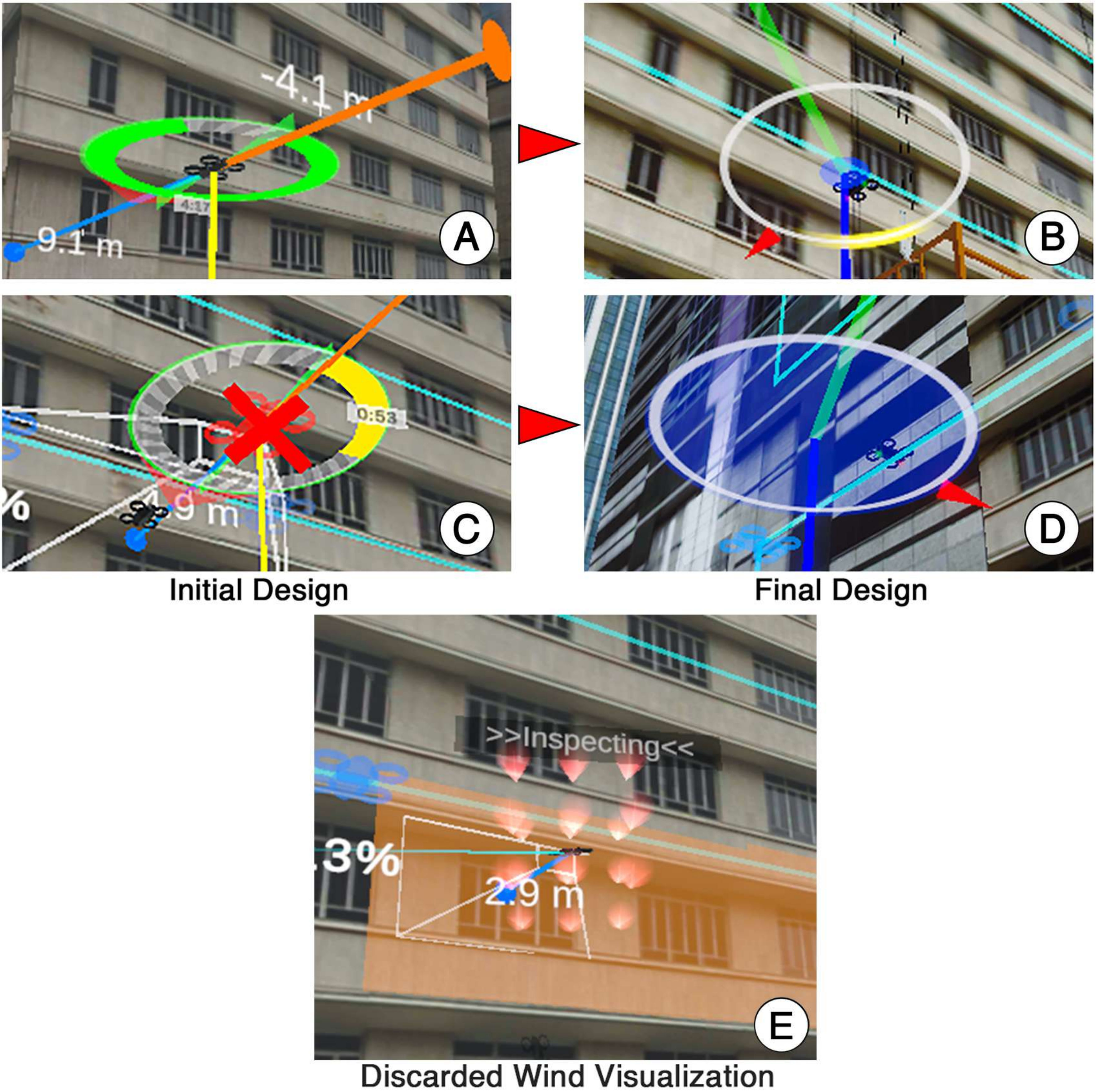}
    \caption{\modif{Iterative Design Decisions}. A \& B: Battery ring, surface projections, and inclination are removed and replaced with a white locator ring, red arrow heading, and circular collision detection. C \& D: GPS error which blocks the real drone is replaced with a more subtle and transparent representation of positional uncertainty. \added{E: Wind visualization was discarded in the final version.}}
    \label{fig:comparison}
    \Description{This figure is to compare the initial designs of the AR drone-centered visualization with our final design, and to show some of the discarded visualizations. The top left is the initial design with a battery ring showing the remaining battery percentage, the inclination of the drone, and the projected distance to the surface of the building and the boundary of the flight zone. The improved version is shown at the top middle, which simply shows a white ring for locating the drone and a red arrow pointing toward the forward direction. The collision status is also visualized on the white ring, showing the distance and the direction of the obstacle. On the bottom left is the old GPS denied visual cue as a red cross on top of the drone. We made the visual cue more subtle by using a transparent blue disc to indicate the GPS uncertainty. The discarded wind visualization is shown on the right, where wind direction is indicated through a cluster of arrows with varying colors. }
\end{figure}

When critical issues occur, safety and control information is displayed in situ through 3D visualizations, primarily as drone-centered visual cues. A horizontal locator ring appears at the drone's real-time position, helping the pilot locate the drone when it is far away, as shown in Figure~\ref{fig:comparison}.B. A red arrow on the ring indicates the drone's heading. Collision detection is visualized on the locator ring in a similar manner to the 2D interface, showing both the direction and distance of nearby obstacles, with colors (yellow or red) indicating proximity. \added{We implemented dynamic scaling for the drone-centered visualization as the drone moves away from the pilot. The scale of the radius remains the same when the distance between the drone and the headset is within 25 meters, while it will grow proportionally when it moves beyond the range ($Scale_{radius}= \max(1.0, distance \times 0.04)$). This ensures that the pilot is able to see the visualization even if the drone is far away. }

\begin{figure*}[!h]
    \centering
    \includegraphics[width=1\linewidth]{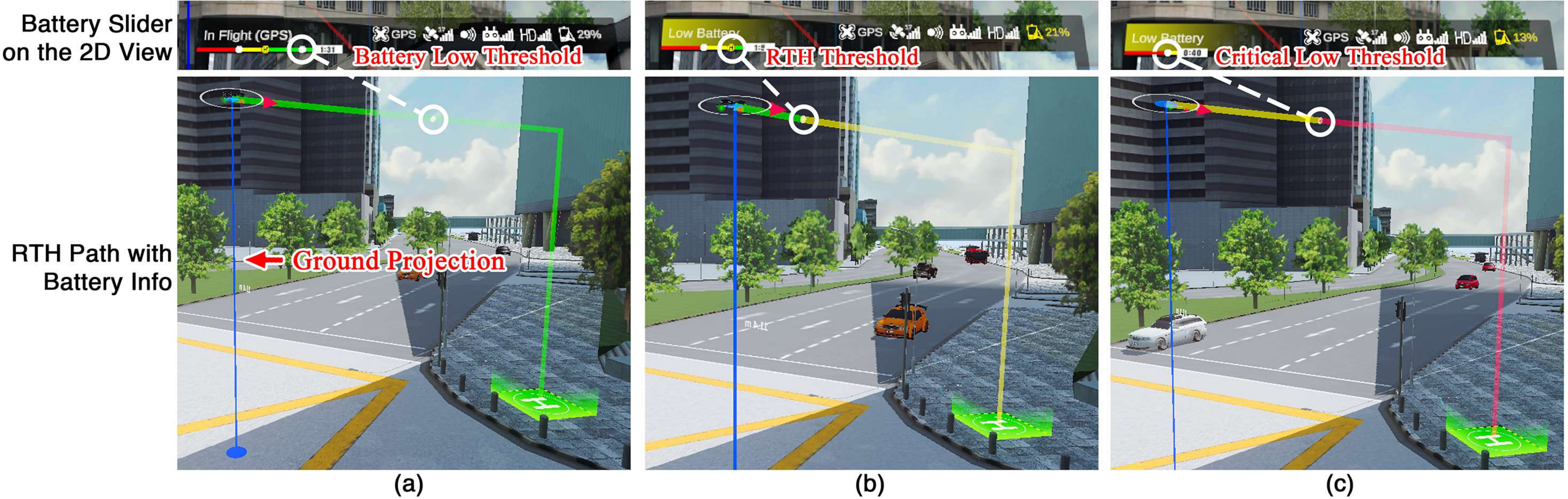}
    \caption{Battery remaining time on the 2D view and the 3D RTH path. (a) The white dot indicates the battery's low threshold (25\% of battery remaining in our case). Ground projection is visualized in blue. (b) The yellow dot indicates the RTH threshold dynamically calculated based on the distance of the drone to the home point. (c) The red dot indicates the battery's critically low threshold (10\% of battery remaining in our case).}
    \label{fig:rth-path}
    \Description{The mechanics of battery remaining time and threshold are explained in this figure. The 2D battery slider and the 3D RTH path are shown in the condition of near battery low threshold, RTH threshold, and the critical low threshold status.}
\end{figure*}

A blue vertical line is projected from the drone's position to the ground level, as presented in Figure~\ref{fig:rth-path}.a. The altitude is shown near the center of the line. This not only visualizes the relative horizontal location of the drone to the surrounding structures, such as the building facade, but also indicates the projected fall area on the ground.

As part of the GPS loss simulation, the accuracy of the drone-centered visualization is not guaranteed during GPS loss. To make the pilot aware of the \added{horizontal} positioning error, the positional uncertainty is visualized as a holographic disc beneath the drone, becoming more transparent as uncertainty increases to avoid blocking the view of the real drone (Figure~\ref{fig:comparison}.D). This allows pilots to notice GPS issues directly from the line of sight and make prompt decisions. In earlier iterations, GPS loss was indicated by a red cross above the drone (Figure~\ref{fig:comparison}.C), but D5 and D6 indicated that it was drawing too much attention away from the actual drone’s position and argued for a more subtle, continuous indicator of positional error.

When the battery is running low, a Return-to-Home (RTH) path is visualized, connecting the drone's current position to the home point as illustrated in Figure~\ref{fig:rth-path}. This path also displays battery sufficiency, similar to the battery slider on the 2D interface, based on suggestions from D1 and D5 during the design workshop. Normally, the path appears green. As the remaining battery approaches the low (25\% of battery left) and RTH (dynamically adjusted based on the distance) threshold, a yellow bar begins to extend from the home point toward the drone's position along the path, indicating the need to trigger the RTH function. When the battery nears the critically low threshold (10\% of battery left), a red bar gradually replaces the yellow, signaling that the drone should land immediately once the entire path turns red.

\subsubsection{Discarded Visualizations}

Several visualizations initially discussed during the design workshops were discarded after pilot feedback. As illustrated in Figure~\ref{fig:comparison}.A, we experimented with displaying battery level on the drone's locator ring, drone inclination, a projected line to the building surface, and the camera frustum. However, pilots found these displays cluttered and unhelpful. D5 remarked that the inclination was \textit{"more disturbing than helpful,"} while D6 and D7 both found the battery ring unnecessary. D7 also suggested simplifying the camera frustum to an arrow to reduce visual clutter, leading us to revise these elements in the final version.

We also tested wind visualization in an early version. \added{We visualized the wind as a cluster of animated 3D arrows pointing towards the detected wind direction. The strength of the wind was indicated by the speed of the animation and the color of the visualization (from white to red, Figure~\ref{fig:comparison}.E).} \modif{However,} D5 and D6 found the cues for wind direction and strength unclear and of little use. In practice, pilots preferred to rely on the camera view and VLOS when adjusting the drone’s position. As a result, we removed the wind indicator to declutter the interface. Additionally, wind and GPS loss zones were not visualized, as they would require additional mission preparation that we did not focus on for our study.

\subsection{Adaptation between Mission and Safety Views}

\begin{figure*}[!ht]
    \centering
    \includegraphics[width=1\linewidth]{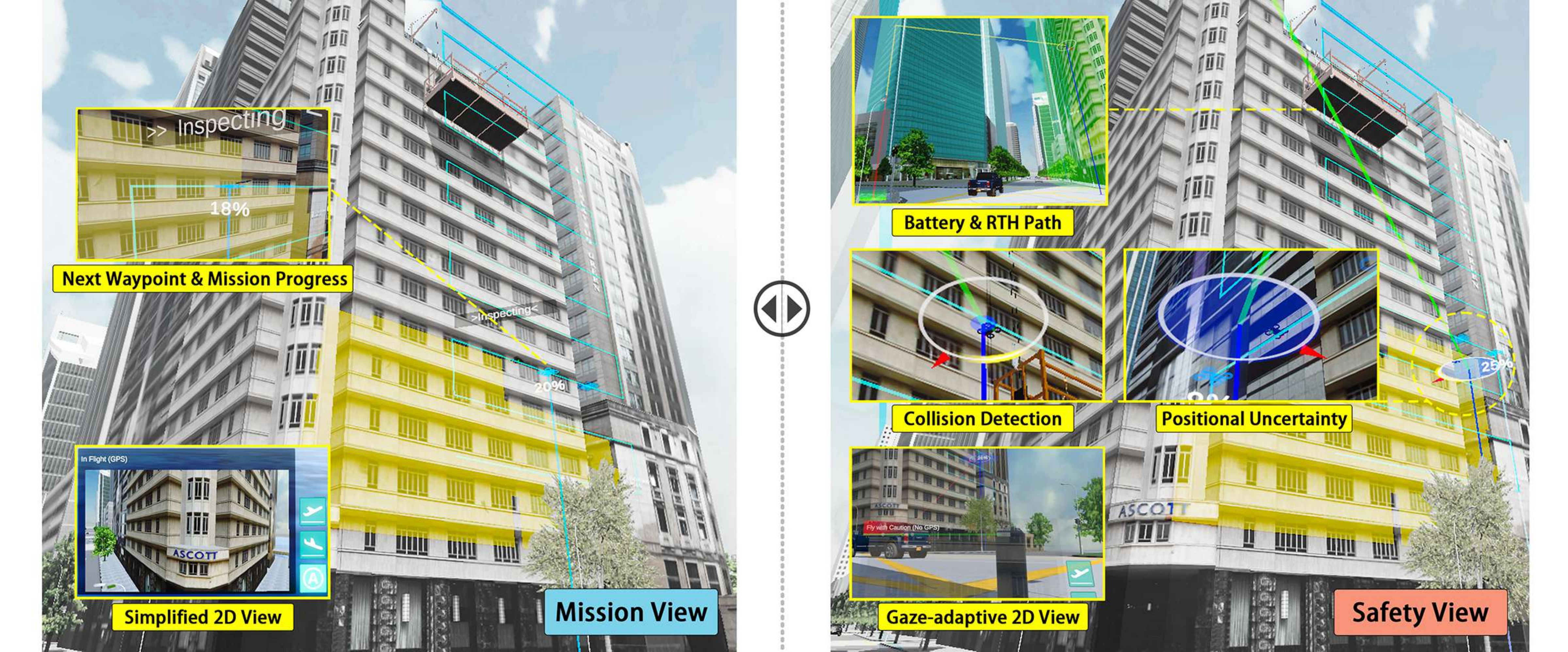}
    \caption{Adaptive behavior during the inspection and the critical phase of the drone flight. Left: Mission view that focuses on marking defects and tracking the mission progress. Right: Safety view that focuses on controlling the drone with both camera view and VLOS and being aware of the safety issues. Note that the colors of the simulated real-world elements have been desaturated to facilitate the visualization of AR elements.}
    \label{fig:adaptive}
    \Description{This is an overview of the adaptive behaviour of the interface during the inspection. The mission view is shown on the left with details of the visualization of the next waypoint and the simplified 2D view. The safety view is shown on the right with details of battery and RTH path, collision detection, positional uncertainty, and the gaze-adaptive 2D view.}
\end{figure*}

Along with pilots, we refined the concept and behavior of the adaptation between two views, mission and safety. Figure~\ref{fig:teaser} and Figure~\ref{fig:adaptive} provide an overview of the differences between the two views.

We mostly relied on the feedback provided by the participants to elicit which visualization would be integrated into which view and how to transition between them. Following principles from Deuschel and Scully~\cite{adapt_ui_guidelines}, we opted for preserving the spatial organization of the information and mostly focused on which information was required in each view to make it visible or not.
%

The first is \textbf{mission view}, which focuses on inspection during autopilot. Here, the 2D interface is centered in the pilot's vision (head-locked), with most interface elements hidden, showing only the 2D status messages, the 3D drone path, and coverage information. 

The second is \textbf{safety view}, which focuses on manual piloting and the visibility of safety issues. The 2D interface is positioned in a body-locked view to assist with line-of-sight piloting, and all drone-centered visualizations—such as the position ring, heading, ground projection, and RTH path are displayed. \modif{Additionally, following the recommendation of Pfeuffer \textit{et al.}~\cite{pfeuffer2021artention} on using gaze to navigate multiple layers of information, we implemented dynamic transparency of the 2D interface based on the pilot's eye gaze approximated by the head-tracking feature (eye-tracking is not available on Meta Quest 3). The interface becomes faded when it is not close to the center of the view space, which is determined by a ray casting at the forward direction of the view camera. }



\added{The system automatically switches from the mission view to the safety view when abnormalities or emergency events are detected. These include near collisions, GPS loss, or low battery, as well as a manual override when the control mode is switched from autopilot to manual piloting (triggered by moving the joysticks). Upon switching to the safety view, AR safety visualizations are displayed instantly, and the 2D interface transitions smoothly from a head-locked to a body-locked position with a brief animation. The pilot can return to the mission view by clicking the autopilot button when no safety issues are present.}

%% file: Chapters/5_Comparative_Study.tex
\section{Comparative Study between 2D-Only, Full-AR and Adaptive-AR }

We aimed to evaluate if the adaptive AR interface that adjusts the views based on the task context can reduce cognitive load and enhance situational awareness. 
We conducted a comparative user study of the three interface designs: a regular 2D-only interface (\textsc{2d-only}), a static full-AR interface that displays all information simultaneously (\textsc{full-ar}), and the adaptive AR interface  (\textsc{adapt-ar}).

\subsection{Method}

\subsubsection{Participants}

\modif{We recruited 15 participants (10 male, 5 female, mean age 24.93 years) through university platforms in Singapore and Toulouse. Participants were required to have prior experience with outdoor drone flights, including manual control or autopilot.}

\modif{These amateur pilots are indexed as A1-A15. The study took approximately 1.5 hours per participant, and they were compensated \$20 for their time. The experiment was approved by the IRB of our university. }


\added{In addition, due to limited availability,  three professional drone pilots who had participated in the iterative design sessions (D5-D7) tested the three conditions of the  prototype to provide qualitative feedback. }

\subsubsection{Study Design}
The experiment follows a within-subject design, where each participant experienced the three conditions: \textsc{2d-only}, \textsc{full-ar}, and \textsc{adapt-ar}. The order of the conditions was counterbalanced using a Latin square to cancel any order effects. 

\subsubsection{Task}

Participants performed a drone-based building inspection task similar to the one described in Section~\ref{sec:system-design} in a virtual reality (VR) environment. The participants had to:

\begin{enumerate}
    \item Identify and mark defects on the building facade,
    \item Follow the predefined flight path and minimize deviations unless obstacles were encountered,
    \item Maintain safety by avoiding collisions and monitoring battery life for a safe landing.
\end{enumerate} 

Three experimental task configurations were created to prevent learning from one experimental condition to the other. Each configuration involved inspecting a different building facade (short or long) with different layouts of defects, mid-air obstacles, and GPS-denied zones. Battery life was set according to the facade size (4 minutes 20 seconds for short, 5 minutes 10 seconds for long). Figure~\ref{fig:config-2d-only}.a shows one of the configurations used in the experiment on the short facade. The three configurations were designed to be almost equivalent in terms of difficulty.

\subsubsection{Conditions}
\begin{figure}[ht]
    \centering
    \includegraphics[width=1\linewidth]{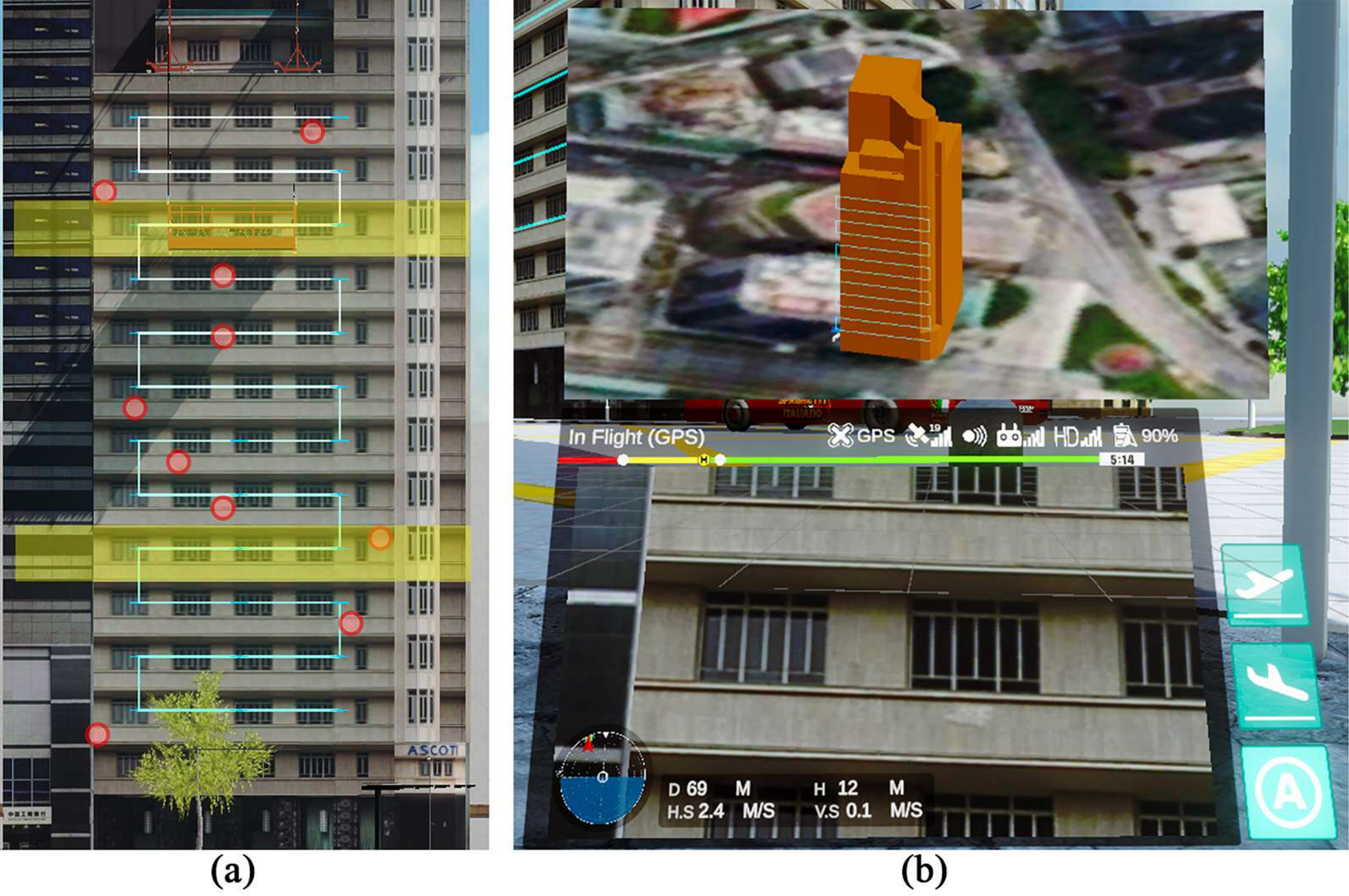}
    \caption{(a) One example of the configuration on the short facade. Yellow regions indicate GPS-denied areas. Red circles surround the defects. (b) 2D tablet-based Interface in the \textsc{2d-only} condition. Top: Corresponding path plan for the config. Bottom: Camera view (DJI GO 4~\cite{DJI_GO_4})}
    \label{fig:config-2d-only}
    \Description{Two figures showing an example of the configuration on the short facade and the simulated 2D hand-held interface. On the left, the position of the GPS-denied zone and the defects are shown to demonstrate how we configure the experiment in one of the configurations. On the right, a conventional 2D drone control interface is displayed at the bottom and a secondary screen showing the path plan over a virtual building is displayed on the top.}
\end{figure}

We compared three Interface conditions:

\paragraph{\textsc{2d-only}} This condition replicates a regular interface displayed on a tablet, similar to the DJI GO 4 app~\cite{DJI_GO_4}, and serves as a baseline. The interface is attached to the user's left hand in VR and displays a camera view and flight path plan simultaneously (Figure~\ref{fig:config-2d-only}). The display replicates the vertical path plan visualizer and autonomous flight monitor of professional inspection software such as UgCS~\cite{SPH_Engineering_2022}.

\paragraph{\textsc{full-ar}} In this condition, all the information (camera view, flight path, safety indicators) is overlaid on the AR interface without adaptation. The interface remains static, showing all visualizations at once. The 2D display is head-fixed within the user’s line of sight, eliminating the adaptive switching between head-fixed and body-fixed positions.

\paragraph{\textsc{adapt-ar}} The adaptive interface, as described in Section~\ref{sec:system-design}, selectively presents either mission-focused or safety-focused information based on the ongoing task context. It aims to simplify the interface by focusing the user's attention on the most relevant visual elements at any given moment.

\begin{figure*}[!h]
    \centering
    \includegraphics[width=1\linewidth]{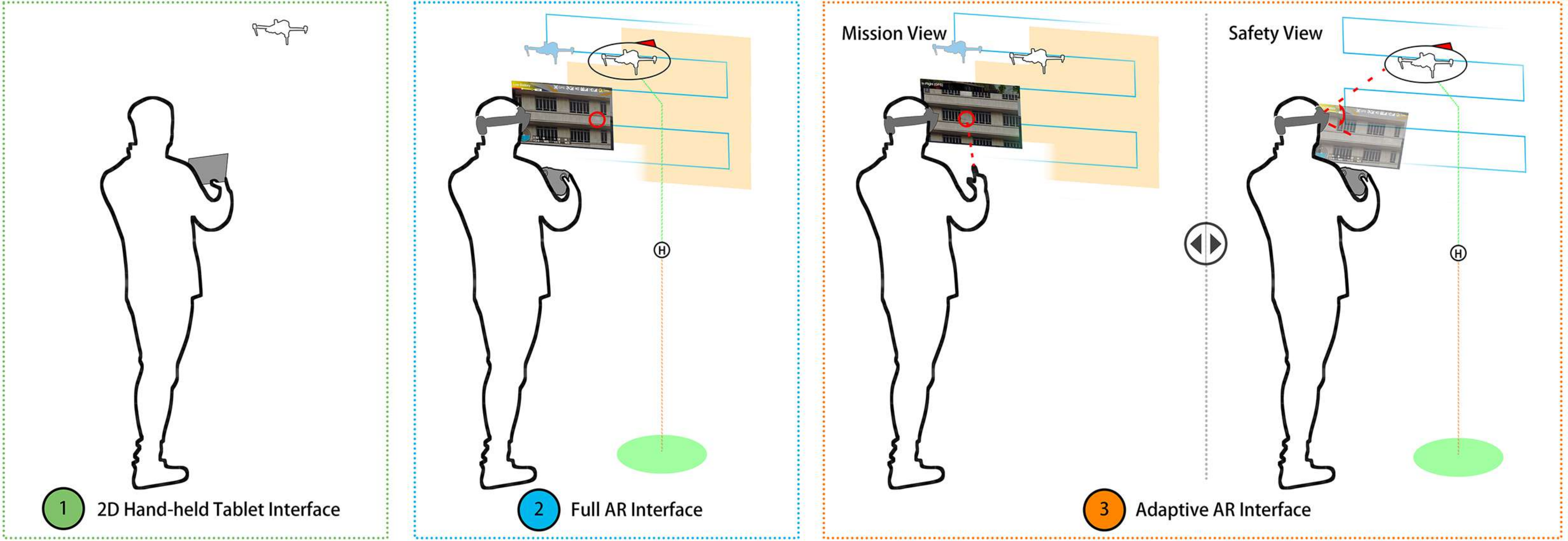}
    \caption{The three interface conditions in the user study. (1) 2D hand-held tablet interface with no AR visualization as the baseline. (2) AR interface that displays all information without adaptive features. (3) Adaptive AR interface that switches between mission and safety views. }
    \label{fig:conditions}
    \Description{An illustration showing the comparison of three interface conditions. On the left, the 2D-only condition is shown with a man holding the conventional 2D tablet drone control interface. In the middle, the AR full visualization condition is displayed with a floating 2D camera view in front of the pilot and all AR visual cues over the building surface and on top of the drone. On the right, the AR adaptive interface is illustrated with a mission view focusing on inspection with the 2D camera view and a safety view focusing on visualizing safety issues and controlling the drone.}
\end{figure*}

\subsubsection{Procedure}
We used the Meta Quest 3 headset for the virtual reality simulation. The application was built as a standalone software pre-loaded on the headset. Participants wore the headset and used the controllers to interact. They were asked to stand within a safe space without obstacles during the experimental tasks.

The experiment began with a brief familiarization session, where participants used the 2D-only interface to take off, practice manual control, and experience simulated GPS loss, drifting, and mid-air obstacles. This helped them get comfortable with the VR environment and the controller.

After familiarization, participants completed a learning phase with their first interface condition where they practiced the inspection task and received performance scores (correctly marked defects, average deviation from the flight path). They could retry the learning phase if unsatisfied with their performance. Following this, participants completed the inspection task under the same interface conditions. During the actual test, the performance scores were hidden.

After each condition, participants completed a post-task questionnaire to evaluate cognitive load, situational awareness, task performance, safety information accessibility, and self-confidence. Finally, one researcher conducted a semi-structured interview to gather qualitative feedback and identify future design opportunities.


\subsection{Data Collection and Analysis}


\subsubsection{Performance}
To assess the impact of the Interface on the task performance, we logged several data and computed two key performance indicators:

\begin{itemize}
    \item Percentage of marked defects: The number of correctly identified and marked defects.
    \item Deviation from flight path: The average shortest distance from each waypoint to the actual navigation path, indicating how well the pilot can follow the flight plan.
\end{itemize}

\subsubsection{Questionnaires}

After each condition, participants were asked to fill out three specific questionnaires \added{(provided in Appendix~\ref{appendix:questionnaire}.}) focusing on different aspects:
\begin{itemize}
    \item Cognitive Load: We used the Bedford Workload Scale to assess cognitive load, a method used by NASA to evaluate cognitive workload~\cite{roscoe1990subjective}. The scale uses a hierarchical decision tree to quantify cognitive load.
    \item Situational Awareness: \added{Since our task are characterized by its dynamic and undefined nature, }we measured situational awareness using the SART (Situational Awareness Rating Technique)~\cite{endsley2000situation} \added{following the guideline of~\cite{salmon2009measuring}}. It evaluates task demand (D), understanding (U), and supply (S) with a final situation awareness score calculated using the formula: SA = U - (D - S).
    \item Participants rated their performance, the accessibility of safety information, and their confidence in using the interface for real-world inspections using a 7-point Likert scale.
\end{itemize}

For both performance and questionnaire data, we used a Repeated Measures (RM) ANOVA test to analyze the data if the normality assumption was met (Shapiro-Wilk test, $p > 0.05$), while for non-parametric data, we used the Friedman test to analyze the main effect and applied Conover's Post Hoc Comparisons. JASP~\cite{JASP2024} was used for data analysis. 

\subsubsection{Semi-structured Interview}

Interviews were recorded and transcribed. One of the authors did a thematic analysis using QDA Miner software~\cite{qda-miner} to identify codes and group them into themes. These themes were consolidated with the other authors.

\subsection{Results}
 
\subsubsection{Quantitative Data}


\paragraph{Task Performances}

The mean percentage of marked defects for \textsc{2d-only} is \modif{83.2}\%, while \textsc{full-ar} is \modif{79.3}\%, and \textsc{adapt-ar} is \modif{71.7}\%. \added{The Shapiro-Wilk test rejects the normality of the data ($p_{adapt-ar}=0.038$). The Friedman test shows no main effect of the interface on the marked percentage ($\chi^2_F=3.000$, $p=0.223$, ${Kendall's\ W}=0.100$).} 


The mean deviation from the flight path for \textsc{2d-only} is \modif{2.84} meters, while \textsc{full-ar} is \modif{2.84} meters, and \textsc{adapt-ar} is \modif{2.62} meters. The Shapiro-Wilk test reveals that the normality is not met for \textsc{2d-only} and \textsc{full-ar} (\modif{$p_{2d-only}=0.004$, $p_{full-ar}=0.002$}). The Friedman test shows no main effect of the Interface on the average deviation (\modif{$\chi^2_F=0.533$, $p=0.766$, ${Kendall's\ W}=0.018$}).

\begin{figure}[!h]
    \centering
    \includegraphics[width=1\linewidth]{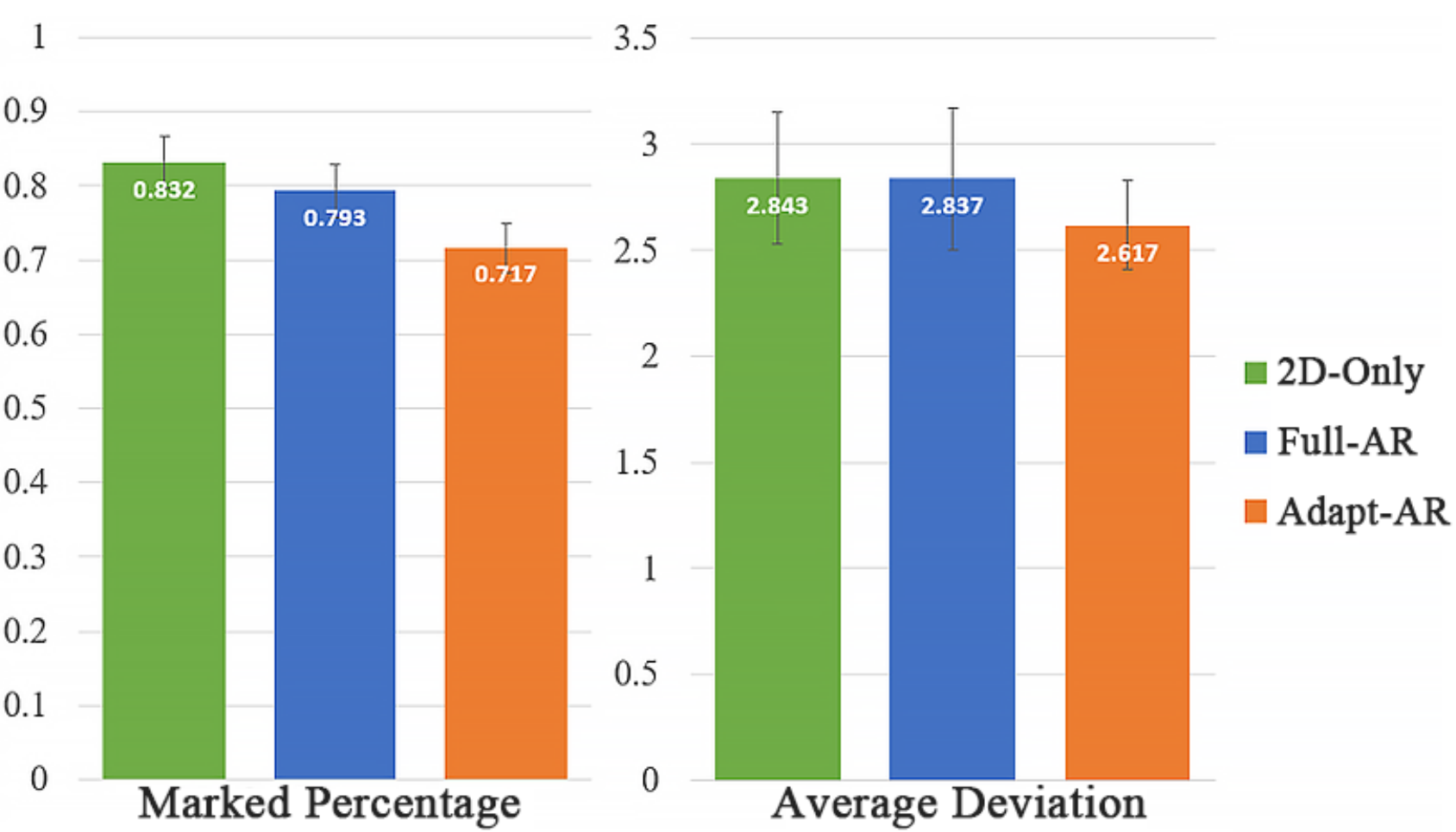}
    \caption{\modif{Mean rating for the percentage of marked defects and average deviation from the flight path.}}
    \label{fig:chart-performance}
    \Description{This is a bar chart showing the mean values of the marked defect percentage and the average deviation from the flight path in all three interface conditions. No significant difference was observed.}
\end{figure}

\begin{figure*}[!h]
    \centering
    \includegraphics[width=1\linewidth]{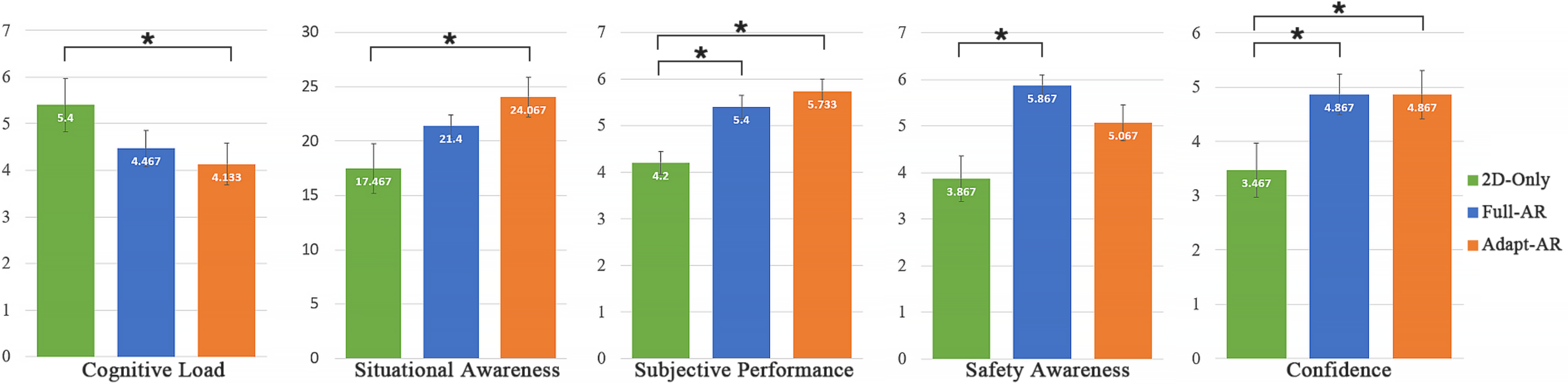}
    \caption{\modif{Mean rating for cognitive load, situational awareness, subjective perception of the performance, accessibility of the safety information, and the confidence of using the current interface. The significant comparisons are highlighted with asterisk marks.}}
    \label{fig:chart}
    \Description{This is a bar chart showing the mean values of the rating and the standard error of cognitive load, situational awareness, subjective perception of the performance, accessibility of the safety information, and the confidence of using the current interface in all three interface conditions. The significant comparisons are highlighted.}
\end{figure*}

\paragraph{Cognitive Load}

The mean cognitive load ratings of \textsc{adapt-ar} is \modif{4.13} (\modif{$SD=1.727$}), while \textsc{2d-only} is \modif{5.40} (\modif{$SD=2.230$}) and \textsc{full-ar} is \modif{4.47} (\modif{$SD=1.506$}).

The Shapiro-Wilk test of \textsc{2d-only} \added{and \textsc{full-ar} }rejected the normality of the data (\modif{$p_{2d-only}=0.060$, $p_{full-ar}=0.008$}). The Friedman non-parametric test shows a significant main effect of the Interface on cognitive load (\modif{$\chi^2_F= 7.581$, $p=0.023$, ${Kendall's\ W}=0.253$}). The Conover's post-hoc tests showed that there is a significant reduction of cognitive load between \textsc{2d-only} and \textsc{adapt-ar} (\modif{$p=0.005$, $p_{bonf} = 0.016$, $p_{holm} = 0.016$}). No significant difference was observed between other pair-wise comparisons. This indicates that the adaptive interface significantly reduces the cognitive load compared to the baseline (the current 2D-only tablet interface), while the full AR visualization has little improvement in terms of cognitive load. 


\paragraph{Situational Awareness}
The mean situational awareness is \modif{17.47} (\modif{$SD=8.879$}) for \textsc{2d-only}, \modif{21.40} (\modif{$SD=4.014$}) for \textsc{full-ar}, and \modif{24.07} (\modif{$SD=7.126$}) for \textsc{adapt-ar}.

As the data passed the Shapiro-Wilk normality test, we did an RM ANOVA test, which shows a significant main effect of the Interface on the situational awareness (\modif{$F(1,14)=5.107$, $p=0.013$, $\eta_{p}^{2} = 0.267$}). The Post hoc test shows a significant difference between \textsc{2d-only} and \textsc{adapt-ar} (\modif{$p_{bonf}=0.021$, $p_{holm}=0.021$}). No significant difference is observed in the other pair-wise comparisons. These results suggest that the adaptive interface improves the pilot' situational awareness during the flight compared to the baseline. 

\paragraph{Subjective Performance}
The subjective performance is rated on a 1-7 scale. The mean subjective performance of \textsc{2d-only} is \modif{4.20} (\modif{$SD=0.941$}) while \textsc{full-ar} is \modif{5.40} (\modif{$SD=0.986$}) and \textsc{adapt-ar} is \modif{5.73} (\modif{$SD=1.033$}). 

The Shapiro-Wilk test shows that normality does not meet for both the data in \textsc{2d-only} and \textsc{adapt-ar} (\modif{$p_{2d-only}=0.004$, $p_{adapt}=0.005$}). The Friedman test shows that there is a significant main effect of the Interface on the subjective performance (\modif{$\chi^2_F=12.360$, $p=0.002$, ${Kendall's\ W}=0.412$}). The Conovor's post hoc test shows significant differences for \textsc{2d-only} vs. \textsc{adapt-ar} (\modif{$p<.001$, $p_{bonf}<.001$, $p_{holm}<.001$}), as well as \textsc{2d-only} vs. \textsc{full-ar} (\modif{$p=0.002$, $p_{bonf} = 0.006$, $p_{holm} = 0.004$}). This indicates that the AR, in general, improves the perceived performance. 

\paragraph{Safety Accessibility}
The accessibility of the safety information is rated on a 1-7 scale. The mean safety accessibility is \modif{3.87} for \textsc{2d-only} (\modif{$SD=1.885$}), \modif{5.87} for \textsc{full-ar} (\modif{$SD=0.915$}), and 5.07 for \textsc{adapt-ar} (\modif{$SD=1.486$}). 


\modif{The Shapiro-Wilk test shows that the normality is met. The RM ANOVA test shows that there is a significant main effect of the Interface on the safety accessibility ($F(1,14)=7.222$, $p=0.003$, $\eta_{p}^{2} = 0.340$). The Conovor's post hoc test shows a significant contrast between \textsc{2d-only} and \textsc{full-ar} ($p_{bonf} = 0.014$, $p_{holm} = 0.014$). This indicates that only the full visualization improves the accessibility of the safety information.}

\paragraph{Confidence}

The confidence of using the interface is rated on a 1-7 scale. The mean confidence is \modif{3.47} for \textsc{2d-only}  (\modif{$SD=1.922$}), \modif{4.87} for \textsc{full-ar} (\modif{$SD=1.457$}), and \modif{4.87} for \textsc{adapt-ar} (\modif{$SD=1.727$}). 

The Shapiro-Wilk test shows that the normality does not meet for the data in \textsc{full-ar} (\modif{$p=0.041$}). The Friedman test shows that there is a significant main effect of the Interface on the confidence (\modif{$\chi^2_F=9.733$, $p=0.008$, ${Kendall's\ W}=0.324$}). The Conovor's post hoc test shows significant differences between \textsc{2d-only} and \textsc{adapt-ar} (\modif{$p=0.006$, $p_{bonf} = 0.018$, $p_{holm} = 0.012$}) and between \textsc{2d-only} and \textsc{full-ar} (\modif{$p=0.002$, $p_{bonf} = 0.007$, $p_{holm} = 0.007$}). This indicates that  AR, in general, improves the confidence of using the interface.

\subsubsection{Qualitative Data}
Participants provided rich feedback on the three interfaces. We present our findings below according to four major themes that emerged during the thematic analysis.

\paragraph{AR Serves as an Assistive Feature for the Mission}

Participants found AR visualization highly beneficial for mission completion. \modif{Many favored the 3D path, with A12 noting, \textit{"I can follow the path instead of flying aimlessly,"} and D6 emphasizing the usefulness of the camera coverage indicators over the facade in tracking progress: \textit{"This one (AR) is definitely helpful... it’s more mission related."} \added{A13} and D6 suggested adding coverage indicators to both the building surface and the camera view for better utility.}

\modif{The Return To Home (RTH) path, showing remaining battery life, was praised by most participants (11/15) for improving safety and flight planning. A2 appreciated knowing when to return, saying \textit{"I know when I need to come back and I will plan (the rest of) the flight accordingly"}.
D6 commended the design as straightforward and informative: \textit{"It tells you how critical your battery life is... and the path is a straight line, which is very useful."}} \added{A15 noted that the feature became increasingly helpful with familiarity, especially in locating power information.}


\paragraph{AR Supports Attention Switches, but Visibility Remains an Issue.}
\modif{The 2D interface posed challenges in maintaining Visual Line of Sight (VLOS) and assessing drone orientation. AR visualizations, such as the drone locator, heading, and ground projection, aided most pilots (11/15) during attention switches. A3 noted, \textit{""You don’t see where the drone really is. Then you can use the drone locator,"} while A6 emphasized its utility when the drone is distant and barely visible. The ground projection also helped pilots assess the drone's position relative to obstacles, as A10 highlighted: \textit{""It has height and position information... relative to the trees."}
The AR collision indicator was useful for some (A1, \added{A13, A14}) but less effective for others, like A6 and D6, who preferred collision indicators in the video feed, as seen in commercial systems. D6 remarked, \textit{"The ring around the drone is not so noticeable... it’s not really catching my attention,"} suggesting that AR displays could be improved for better visibility near obstacles, especially when the drone is far from the pilot.
}

\paragraph{Adaptive Visualization Potentially Reduces Cognitive Load, but Requires Trust in the System.}

The AR visualization can help the pilot with mission tracking and attention-switching, but the AR interface with full visualizations provides little improvement in cognitive load compared to the baseline 2D interface. However, both quantitative and qualitative feedback indicate that the adaptive interface significantly reduces the cognitive load over the baseline. \modif{D5} mentioned that \textit{"adaptive was better because it hides complexity and reduces the cognitive workload during automated flight phases."}. A7 said that it felt \textit{"less stressful"} when using the adaptive mode. 

\modif{Participants raised concerns about trusting the adaptive system. D6 preferred having all critical information visible at all times, stating,  that \textit{"When I want it to appear, it doesn't appear and then it becomes a bit stressful."} A10 \added{and A15} shared similar sentiments with \modif{A10} noting, \textit{"if everything is shown, I can focus on what I need and ignore the rest... if you don't show it to me, I feel insecure."} D5 had a more balanced perspective, initially feeling \textit{"surprised"} by the lack of information but later appreciating the reduced clutter for improved focus. However, he added, \textit{"You have to trust the system to actually detect problems."}} This lack of trust contributed to a lower safety information accessibility rating for the adaptive interface compared to the full AR interface.

\paragraph{Adaptive Visualization Makes Warning more Salient, but Screen Placement and Behavior Need Improvement.}

\modif{Participants found the adaptive interface effective in providing clear warning signals when switching to the safety view, helping them react quickly. D5 noted, \textit{"When I use autopilot, I need less concentration because I can use the alarms to react... it's even more important in adaptive as the camera view changes to make you realize there’s a problem."} A9 found the adaptive visualization helpful during manual piloting by removing irrelevant information, and A4 and \added{A13 appreciated how it uncluttered the visual line of sight}.}


\modif{However, the inability to adjust the location and scale of the 2D interface in AR hindered usability, particularly for the adaptive interface. A7 noted discomfort with the 2D interface \textit{"following my head"}, while D6 and A11 preferred the flexibility of a hand-attached 2D interface. D5 also wished to remove the 2D interface from view, explaining, \textit{"The screen is always in the field of view. It forces you to look at it."} \added{ Additionally, A13 mentioned a lag when the 2D interface repositioned, indicating that the head-tracking feature was not responsive enough, reducing overall usability.}}

%% file: Chapters/6_Discussion_Conclusion_Future_Work.tex
\section{Discussion}

This section synthesizes insights from our design workshops, iterative  process, and user study to address our research questions and highlight key design considerations for improving human-drone interaction with heads-up AR displays. 

\subsection{RQ1: Design Considerations for Reducing Cognitive Load in a Heads-up AR Drone Control Interface}

Our results show that an adaptive AR interface can effectively decrease cognitive load. \modif{However, the rating (3.67 as visible in Figure~\ref{fig:chart}) still slightly exceeds NASA's recommended threshold for nominal tasks~\cite{nasa_cog_load} (a score of 3 or below)}. This may be due to participants' unfamiliarity with the interface compared to the more conventional 2D-only interface. With longer use and then higher familiarity, the workload rating could improve~\cite{tuovinen1999comparison}. Additionally, allowing users to customize and control the adaptive interface could further optimize cognitive load and enhance user experience~\cite{10.1145/1166253.1166301}. 
Below, we describe key design considerations based on our findings.

\paragraph{Optimizing Attention Switches with Adaptive AR}

Drone control interfaces are inherently complex, requiring pilots to monitor multiple pieces of information simultaneously. 
Situational impairments such as vision tunneling, external distraction, or screen reflection can exacerbate these challenges and result in safety hazards~\cite{borowik_mutable_2022, rahmani_working_2023}.


Heads-up AR devices, such as HoloLens 2, provide a wider display space that can alleviate information clutter observed in tablet-based interfaces. By appropriately presenting critical information and hiding non-essential data~\cite{Mayer_2009}, AR designers can reduce cognitive load. For example, when GPS is lost, AR could prominently display the drone locator and heading information while hiding less critical data such as the building coverage. This approach can also be applied during automated phases to focus on the main inspection task or any other critical events by identifying relevant data in each situation. In summary, \textbf{with adaptive AR interfaces, designers could reduce complexity and enhance attention-switching efficiency, particularly in critical scenarios}.

    
\paragraph{Enhancing Visibility and Informativeness}

\modif{Visibility and informativeness of AR cues are crucial for cognitive load and situational awareness. \added{In our study, while we implemented dynamic scaling for the drone-centered visualization as it moved away from the pilot, some pilots found these cues most useful when the drone was nearby, with effectiveness decreasing as the drone moved further away from the building.} The heading visualization also lost its usefulness with distance. Additionally, collision detection was not prominent enough and was often missed in AR due to visual cues being too far from the pilot.
\added{At the same time, AR visualizations should avoid obstructing the real-world view or cluttering the space.} Since the user's gaze is approximated by head-tracking, future designs could \textbf{improve clarity by increasing size, enhancing contrast, or using eye-tracking to emphasize key elements}.}

Moreover, visual cues should guide the pilot’s actions. For example, the RTH path with battery status provided clear guidance on when to land. While the AR inspection path and waypoints were useful, participants struggled to align with the path manually due to visual ambiguities. One participant suggested adding an offset indicator to aid alignment during manual piloting. Thus, \textbf{AR visual cues must be informative and adaptable to support decision-making without overwhelming users}.

\paragraph{Streamlining Operations with Dual Control Mode}

Current commercial drones often interrupt autopilot when issues such as GPS loss or wind turbulence are detected. 
Our study suggests that AR can enhance the perception of mode interruptions and facilitate transitions between control modes. However, as indicated by several pilots and our observations, pilots often failed to resume to mission mode when safety issues were resolved. 
Adding clear indications about when to take manual control and how to switch back to autopilot if conditions are stable could help pilots transiting between modes. Also providing specific information on how to re-attend to the mission tasks after focusing on safety could be helpful. For instance, participants suggested adding visual cues on the future direction of the drone when the autopilot will be resumed. Future AR interfaces should \textbf{clearly delineate signals and actions required for transitioning between manual and autopilot modes to streamline the semi-autonomous drone operation}.

\paragraph{Customizing and Controlling the Adaptive AR Interface}

Our work revealed diverse preferences for the placement and visualization of specific visual elements, both in AR and on the 2D interface. 
Unlike traditional tablets, heads-up displays offer flexibility in adjusting scale, transparency, and position. It also offers the possibility of multimodal interactions with gestures, voice, and gaze input capabilities~\cite{nizam2018review} even if voice command could be better in piloting tasks compared to gesture-based interaction \cite{li2022influence}. On the other hand, tablets can also be put away to rapidly remove interface clutter.
Users could benefit from interaction allowing them to tailor the interface according to their preferences, the specific flight conditions or their mission context both before the operation or with explicit control while operating. Future designs should \textbf{include customizable options for interface placement, content, and interaction modes, enhancing user control and satisfaction}.



\subsection{RQ2: Safety Considerations for Dealing with Critical Scenarios Using a Heads-up AR Interface}

AR visualization generally enhances perceived safety by improving situational awareness and depth perception~\cite{zollmann_flyar_2014}. In our study, participants gave better safety ratings to the non-adaptive AR visualization even if they rated the adaptive version with a better situation awareness score. 
According to Lavie \textit{et al.}~\cite{LAVIE2010508} high levels of adaptivity support performances in routine situations but could impair performances in unfamiliar situations. 
Based on our results, we discuss below design considerations for improving safety with an adaptive interface without compromising the existing advantages.

\paragraph{Building Trust with the System}

Trust is crucial for adaptive systems that automatically adjust information visibility. Users may initially feel insecure when critical data is hidden. Trust is enabled by the "predictability of the behavior of other agents through understanding their motivations"~\cite{Abbass2018}. The "transparency" of the autonomous agent system is also seen as an important factor for human trust~\cite{lyons2013being}. 
To build trust, we thus suggest to \textbf{enhance transparency by explaining the rules and triggers behind adaptive behaviors, such as why certain information is displayed or hidden}. For instance, the interface can notify users when battery levels necessitate immediate RTH actions and explain the dynamic calculation of thresholds.

\paragraph{Improving Warning Saliency}

The adaptive interface’s ability to highlight warnings was found valuable, as removing the 2D screen forced pilots to look at the drone, but participants suggested stronger notifications than the ones we used. Participants made several suggestions such as visualizing the entire mission path with different colors to indicate the GPS loss or using audio alarms as in some commercial systems they were using. According to Woodward \textit{et al.}~\cite{woodward2023designing}, primary information must be in the field of view to capture attention. For secondary information, having world-anchored elements helps users by letting them decide when to look at the information without distracting them.
Future AR applications for drone piloting should \textbf{enhance the saliency of safety warnings in the AR interface by using distinct visual or auditory cues and positioning critical information close to the drone or in their field of view, but without obstructing the direct view of the drone}. This approach ensures that pilots can promptly recognize and respond to safety issues.

\subsection{Threats to Validity}

\modif{Most professional pilots in our study are not committed in building inspections, with only one (D7) having experience. However, participatory design workshops and iterative design phases included pilots with over 15 years of experience, ensuring the validity of our findings. For the comparative study, we recruited amateur pilots with drone expertise, such as extensive use for photography, test flights, or competitions, as professional pilots were difficult to recruit. Future research could benefit from a more diverse participant pool to better compare professional and recreational pilots.}

\added{While our study gave significant differences for subjective performance measures, the objective performance measures showed no significant difference. Using G*Power 3.1, we calculated the desired sample size for the marked percentage based on our existing data's effect size, which indicated a sample of 11. With 15 participants, the lack of significance is likely due to non-normal data distribution, limiting the sensitivity of the ANOVA test. Increasing the number of participants may yield significant differences. However, our primary goal was to examine the effect of our design on five factors (cognitive load, Situational Awareness, Subjective Performance, Safety Awareness, and Confidence) in a realistic building inspection scenario.
Regarding Situational Awareness we chose the SART questionnaire as it is quick and easy to administrate after the task but it has been found to correlate with performances and not only situational awareness~\cite{salmon2009measuring}. Using freezing techniques such as SAGAT was not possible in our case since it would interrupt the task and lead to degraded pilot SA and additional efforts to restart.
Future studies could focus on the effect of adaptive AR interfaces on piloting performance more specifically and leverage specific during-the-task questions to assess Situational awareness.}

\added{Our study was conducted in a simulated VR environment but real-world AR conditions pose additional challenges, such as adapting to varying lighting conditions that affect rendering fidelity. Optical see-through devices like HoloLens 2 struggle in bright light, and video see-through systems like Meta Quest 3 require sufficient ambient light. Furthermore, accurate positioning and tracking of virtual objects in our study relied on technologies like Real-Time Kinematics~\cite{de2023systematic} and computer vision~\cite{Mithun_2023}, though environmental changes can cause drift, impacting results. These challenges highlight the need for further research to address real-world discrepancies in AR applications.}

\section{Conclusion}

This paper explores the potential of adaptive AR interfaces to support safety and efficiency in drone facade inspections. We addressed the challenges of managing pilot information overload and using AR in safety-critical situations. Through participatory design workshops and iterative testing with professional pilots, we developed an adaptive AR system that prioritizes information based on task urgency. This system was implemented and evaluated in a VR drone inspection simulator, comparing it against non-adaptive AR and traditional 2D interfaces.

The results reveal that the adaptive AR interface significantly reduces cognitive load and improves situational awareness compared to traditional 2D interfaces and non-adaptive AR interfaces. Key findings from our study suggest that incorporating adaptive design elements can optimize cognitive load by reducing attention switches, enhancing visibility and informativeness, and streamlining the semi-autonomous operation. 

Our findings highlight key factors for improving the safety support of AR interfaces. Enhancing the transparency and explainability of the AR system’s adaptive behaviors can increase pilot confidence. Additionally, making safety warnings and adaptive behaviors more salient through distinct visual cues ensures that pilots can quickly recognize and respond to potential hazards.

We believe that the principles of adaptive AR interface design described in this paper could be applied to other drone-related areas, such as search and rescue missions, traffic management, and human-swarm interaction, where managing information overload and enhancing safety are equally critical.

Future work will explore customization options and interactions that allow pilots to control adaptive behaviors, addressing both trust and performance requirements. Additionally, we aim to validate the effectiveness of adaptive AR interfaces in real-world scenarios.


%% file: Chapters/7_Appendix.tex
\section{Post-Study Questionnaire} \label{appendix:questionnaire}

\begin{figure}[!h]
    \centering
    \includegraphics[width=1\linewidth]{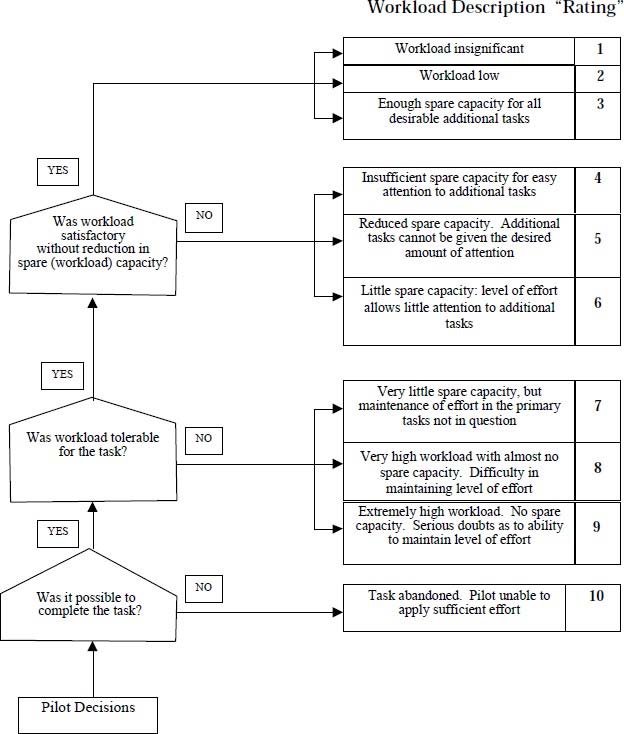}
    \caption{\added{Bedford Workload Rating (adapted from~\cite{theodore2014effect}).}}
    \label{fig:bedford-workload}
    \Description{This is a flowchart showing the Bedford Workload Rating. The participant starts from the bottom of the flowchart and make decision through a tree-like structure to arrive at the final score. }
\end{figure}

\paragraph{Cognitive Load} \added{The Bedford Workload Rating scale used in our study is shown in Figure~\ref{fig:bedford-workload}.}

\paragraph{Situational Awareness} \added{The SART questionnaire consists of 3 different categories of situational awareness and 10 questions in total, each rated on basis of a 7 point Likert scale.}

\begin{enumerate}
    \item Demand
    \begin{itemize}
        \item \added{Instability of Situation: How changable is the situation? Is the situation highly unstable and likely to change suddenly (high) or is it very stable and straightforward (Low)?}
        \item \added{Complexity of Situation: How complicated is the situation? Is it complex with many interrelated components (high) or is it simple and straightforward (low)?}
        \item \added{Variability of Situation: How many variables are changing within the situation? Are there a large number of factors varying (high) or are there very few variables changing (low)?}
    \end{itemize}
    \item Supply
    \begin{itemize}
        \item \added{Arousal: How aroused are you in the situation? Are you alert and ready for activity (high) or do you have a low degree of alertness (low)?} 
        \item \added{Concentration of Attention: How much are you concentrating on the situation? Are you concentrating on many aspects of the situation (high) or focused on only one (low)?}
        \item \added{Division of Attention: How much is your attention divided in the situation? Are you concentrating on many aspects of the situation (high) or focused on only one (low)?}
        \item \added{Spare Mental Capacity: How much mental capacity do you have to spare in the situation? Do you have sufficient to attend to many variables (high) or nothing to spare at all (low)?}
    \end{itemize}
    \item Understanding
    \begin{itemize}
        \item \added{Information Quantity: How much information have you gained about the situation? Have you received and understood a great deal of knowledge (high) or very little (low)?}
        \item \added{Information Quality:  How accessible and usable is the information you have in this situation? Is the knowledge communicated very useful (high) or is it not usable at all (low)?}  
        \item \added{Familiarity with Situation: How familiar are you with the situation? Do you have a great deal of relevant experience (high) or is it a new situation (low)?}
    \end{itemize}
\end{enumerate}

\paragraph{Other Questions} \added{Similarly, these questions were rated on the basis of a 7-point Likert scale.}

\begin{itemize}
    \item \added{How do you rate your performance with the current interface?}
    \item \added{How do you rate the current interface in terms of the accessibility of the safety information?}
    \item \added{How confident are you with using the current interface to conduct building inspection mission by yourself in the real world?}
\end{itemize}